\documentclass[secnumarabic, amssymb, nobibnotes, aps, prb, reprint, superscriptaddress]{revtex4-2}

\usepackage[utf8]{inputenc}
\usepackage[english]{babel} 

\usepackage{amsmath}
\usepackage{amssymb}
\usepackage{upgreek}
\usepackage{gensymb} 
\usepackage{xfrac} 

\usepackage{graphicx}
\usepackage{float}
\usepackage[font=small, justification=raggedright, format=plain]{caption} 

\usepackage{hyperref}

\usepackage{xcolor, soulutf8} 
\begin{document}
        \title{
        Disentangling the Electronic and Lattice Contributions to the Dielectric Response of Photoexcited Bismuth}

	\author{F.~Thiemann}%
	\email[]{fabian.thiemann@uni-due.de}
	\affiliation{Department of Physics, University of Duisburg-Essen, Lotharstrasse 1, 47057 Duisburg, Germany}
	\author{G.~Sciaini}
	\affiliation{The Ultrafast Electron Imaging Lab, Department of Chemistry, and Waterloo Institute for Nanotechnology, University of Waterloo, Waterloo, Ontario N2L 3G1, Canada}
	\author{A.~Kassen}
	\affiliation{Department of Physics, University of Duisburg-Essen, Lotharstrasse 1, 47057 Duisburg, Germany}
	\author{T.S.~Lott}
	\affiliation{The Ultrafast Electron Imaging Lab, Department of Chemistry, and Waterloo Institute for Nanotechnology, University of Waterloo, Waterloo, Ontario N2L 3G1, Canada}
	\author{M.~\surname{Horn-von Hoegen}}
	\affiliation{Department of Physics, University of Duisburg-Essen, Lotharstrasse 1, 47057 Duisburg, Germany}
	\affiliation{Center for Nanointegration (CENIDE), University of Duisburg-Essen, Carl-Benz-Str. 199, 47057 Duisburg, Germany}
	
	\date{June 2022}%
	
	\begin{abstract}
		Elucidating the interplay between nuclear and electronic degrees of freedom that govern the complex dielectric behavior of materials under intense photoexcitation is essential for tailoring optical properties on demand. However, conventional transient reflectivity experiments have been unable to differentiate between real and imaginary components of the dielectric response, omitting crucial electron-lattice interactions. 
		Utilizing thin film interference we unambiguously determined the photoinduced change in complex dielectric function in the Peierls semimetal bismuth and examined its dependence on the excitation density and nuclear motion of the $\mathrm{A_{1g}}$ phonon. Our modeled transient reflectivity data reveals a progressive broadening and redshift of Lorentz oscillators with increasing excitation density and underscores the importance of both, electronic and nuclear coordinates in the renormalization of interband transitions.
	\end{abstract}
	\maketitle

	
        To this day, ultrashort light pulses remain as powerful probes for monitoring a variety of out-of-equilibrium dynamics through sudden changes of the optical dielectric properties within the photoexcited material \cite{Schoenlein1987, Cheng1990}. 
        In particular, a detailed examnination of the dielectric function provides further insights , such as the electron-phonon coupling strength in metals \cite{Obergfell2020} and the identification of exciton properties in transition metal dichalcogenides \cite{Calati2021}. Although these effects are also accessible through other experimental techniques, the main advantage of qualitatively understanding how these affect the optical response is the ability to manipulate the latter by changing electron or lattice degrees of freedom. However, the complexity of the dielectric response escalates as the number of degrees of freedom participating increases. Disentangling these intertwined contributions presents one of the current challenges in ultrafast science \cite{Gerber2017, Otto2018}. In particular, the various degrees of freedom in correlated materials like $\mathrm{VO_2}$, $\mathrm{TaS_2}$, $\mathrm{TiS_2}$, other CDW compounds \cite{Demsar2002, Wall2012} or group V semimetals \cite{Cheng1991} are of broader interest. To target this topic a wealth of transient reflectivity studies has been conducted in the semimetal Bismuth (Bi) \cite{Cheng1990,Hase1996, DeCamp2001,Boschetto2008,Shin2015, Teitelbaum2018}, known for its intrinsic Peierls distortion \cite{Peierls1991}, low carrier density \cite{Liu1995}, and small effective mass \cite{Hofmann2006}. When photoexcited, Bi undergoes a significant increase in carrier density, resulting in a transient modification of the atoms' potential energy surface, which, in turn, launches coherent $\mathrm{A_{1g}}$ phonons \cite{SokolowskiTinten2003, Fritz2007, Sciaini2009}. Despite this body of work, the interconnection between carrier and nuclear dynamics and the dielectric response remains unclear \cite{Timrov2012}.

	Here, we studied the change of the complex dielectric function, $\Delta \varepsilon(\hbar\omega)$, upon impulsive photoexcitation through broadband fs transient reflectivity. Measurements were carried out on a series of (111)-oriented Bi films \cite{Kammler2005,Payer2012} epitaxially grown on Si(111). Film thicknesses, $d$, of 17, 28, 39, 42 and 197\,nm allowed us to exploit interference effects and obtain an unambiguous solution for the lattice and electronic contributions to the transient dielectric function. 
	
	The films were excited by 160-fs optical pulses with a central wavelength of $\lambda =$ 800\,nm (1.55\,eV), providing a maximum absorbed fluence of $0.23\,\mathrm{mJ/cm^2}$. The optical response of Bi was probed by time delayed, $\Delta t$, white light pulses. Transient reflectivity changes $\Delta R/R_0(\hbar\omega, \Delta t)$ in the range of $\lambda =$ 580\,nm to 700\,nm (2.1\,eV to 1.8\,eV) were recorded using a dispersive spectrometer. Ex-situ ellipsometry (see Supplemental Material) was used to verify the thickness of the Bi films and the consistency of the optical properties in comparison with existing studies \cite{Toudert2017}.
	
	
	\begin{figure}[t!]
		\centering\includegraphics[width=1.0\columnwidth]{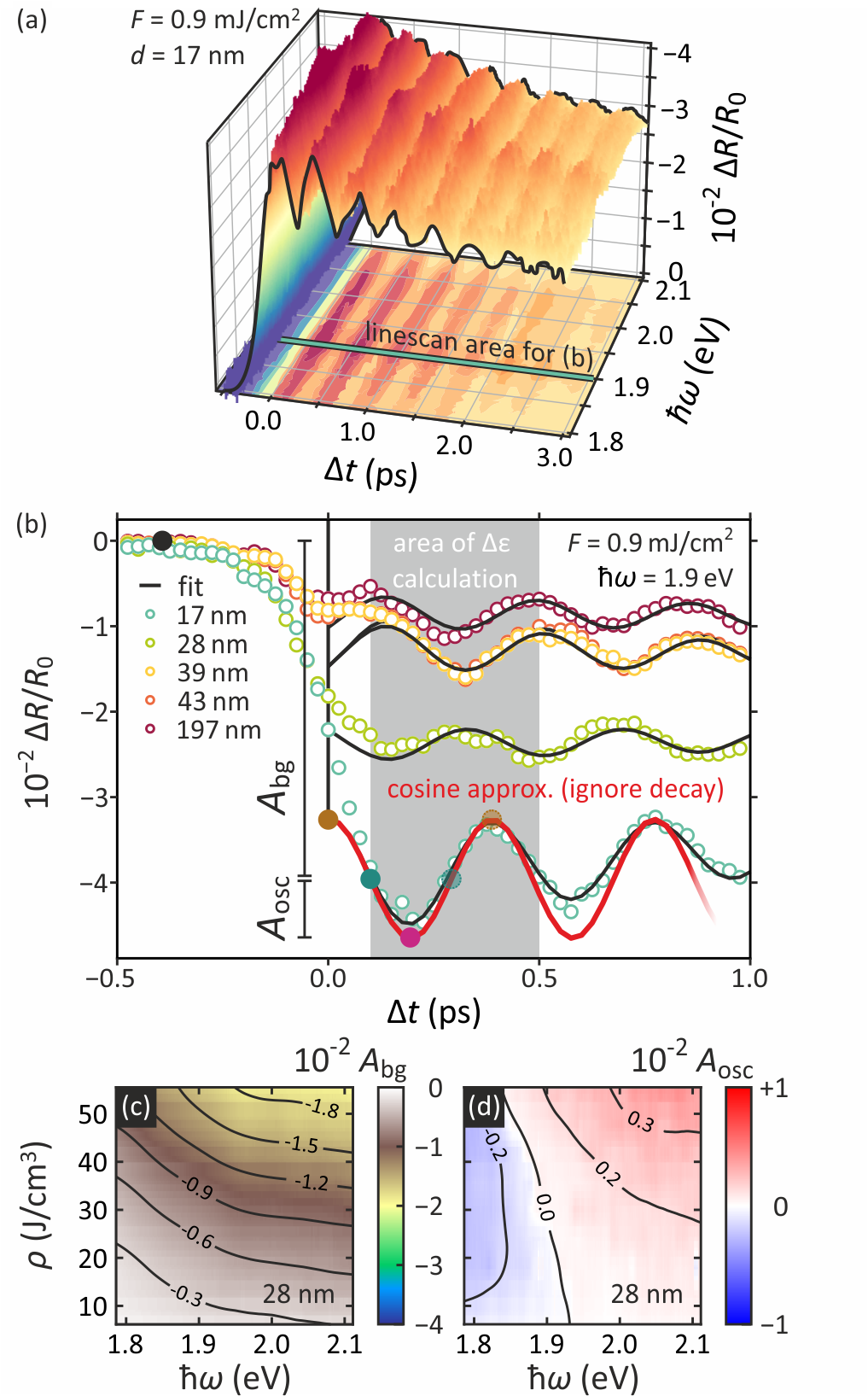}
		\caption{
			(Color online) (a) $\Delta R/R_0 (\Delta t, \hbar \omega)$ for a 17\,nm thick Bi film. (b) Temporal traces, averaged across the teal line in (a), obtained at a photon energy of 1.9\,eV for films of different thickness. The grey area marks where the calculation of  $\Delta \varepsilon$ was conducted. The black lines mark the fits obtained with the model described in Eq. \eqref{eq:model}. The large colored dots refer to distinct points in the potential energy surface of the distorted lattice (see Fig. \ref{fig:two} (a)). (c) and (d) show the amplitudes $A_\mathrm{bg}$ and $A_\mathrm{osc}$, as indicated by the solid black bars in panel (b), for the 28\,nm thick film.
		}
		\label{fig:one}
	\end{figure}
	
	Figure \ref{fig:one} (a) shows a typical time-resolved broadband transient reflectivity spectrum $\Delta R/R_0(\hbar \omega, \Delta t)$ obtained from the 17-nm thick Bi film with characteristics such as the initial sharp jump of reflectivity, the oscillations at $\approx$ 3\,THz, and the decaying background. The spectral dependence exhibits a minimum around 2\,eV. For further analysis, we applied a moving average window with a width of 0.015\,eV (teal line). Temporal traces within $\Delta t$ (-0.5\,ps to 1.0\,ps) are shown in Fig. \ref{fig:one} (b) for five different Bi film thicknesses at a photon energy of $\hbar\omega =$1.9\,eV and an incidence fluence of $0.9\,\mathrm{mJ/cm^2}$. It is noteworthy that the transients exhibit a flip of sign in their oscillation amplitude between 28\,nm~$< d <$~39\,nm.
	
	Since the reflectivity change $\Delta R/R_0(\Delta t,d)$ is the essential input for the calculation of the dielectric function change $\Delta \varepsilon$, we systematically describe it by fitting a phenomenological model \cite{Zeiger1992} for all thicknesses $d$, photon energies $\hbar \omega$ and pump fluences $F$ as follows,
	\begin{equation}\label{eq:model}
		\frac{\Delta R}{R_0}\left(t\right)=A_\mathrm{bg}\, e^{- \frac{t}{\tau_\mathrm{bg}}} + A_\mathrm{osc}\,e^{- \frac{t}{\tau_\mathrm{osc}}} \cos{\left(\Omega t\right)}\,.
	\end{equation}
	The model consists of two terms, the first one describing the decaying background with initial amplitude $A_\mathrm{bg}$ and time constant $\tau_\mathrm{bg}$. The second one, representing the oscillations arising from $\mathrm{A_{1g}}$ phonons, with initial amplitude $A_\mathrm{osc}$, dephasing time constant $\tau_\mathrm{osc}$, and angular frequency $\Omega$. The latter was obtained independently by Fourier analysis and found to change linearly with fluence. The terms are usually referred to as “electronic” and “phononic” \cite{Misochko2004, Ishioka2006, Shin2018}. Such references are not used in this work, because in general both parts include carrier and lattice contributions.
	
	The incident fluence $F$ is an inappropriate parameter to compare the excitation in layers of different thicknesses $d$. In a previous work \cite{Thiemann2022} we demonstrated that ultrafast transport of non-thermalized carriers redistributes the excitation throughout thin films ($d<60\,\mathrm{nm}$) within 150\,fs after excitation. Afterwards the film is homogeneously excited, quantified by the absorbed energy density $\rho = F_\mathrm{abs}/d$, where $d$ is either the film thickness or the effective carrier skin depth $d_\mathrm{eff}$ as introduced in \cite{Jnawali2021} and $F_\mathrm{abs}$ the absorbed fraction of incident fluence. Within the time window of $150\,\mathrm{fs}$ to $500\,\mathrm{fs}$ (grey area in Fig. \ref{fig:one} (b)) and for the sake of a simple analysis we approximate the transients of $\Delta R/R_0 (\Delta t, \rho)$ by a cosine with the amplitude $A_\mathrm{osc}$ and offset $A_\mathrm{bg}$, excluding the decays described by $\tau_\mathrm{bg}$ and $\tau_\mathrm{osc}$. They can be ignored in our analysis, because the cooling \cite{HanischBlicharski2021} and dephasing behavior \cite{Shin2015, He2020} depends on $d$ and the thermalized excited carriers remain at the valence band maximum (\emph{T}-point) and the conduction band minimum (\emph{L}-point) for a few ps prior to recombination \cite{Timrov2012}, respectively. The amplitudes $A_\mathrm{osc}(\rho, \hbar\omega)$ and $A_\mathrm{bg}(\rho, \hbar\omega)$, interpolated to the same $\rho$, are shown in Fig. \ref{fig:one} (c) and (d). The most characteristic pattern of $A_\mathrm{osc} (\rho, \hbar\omega)$, observed for the 28\,nm film, features the flip of sign in the $\rho$-$\hbar \omega$-plane, similar to the one in Fig. \ref{fig:one} (b) for different film thicknesses.
	
	
	\begin{figure}[t!]
		\centering\includegraphics[width=1.0\columnwidth]{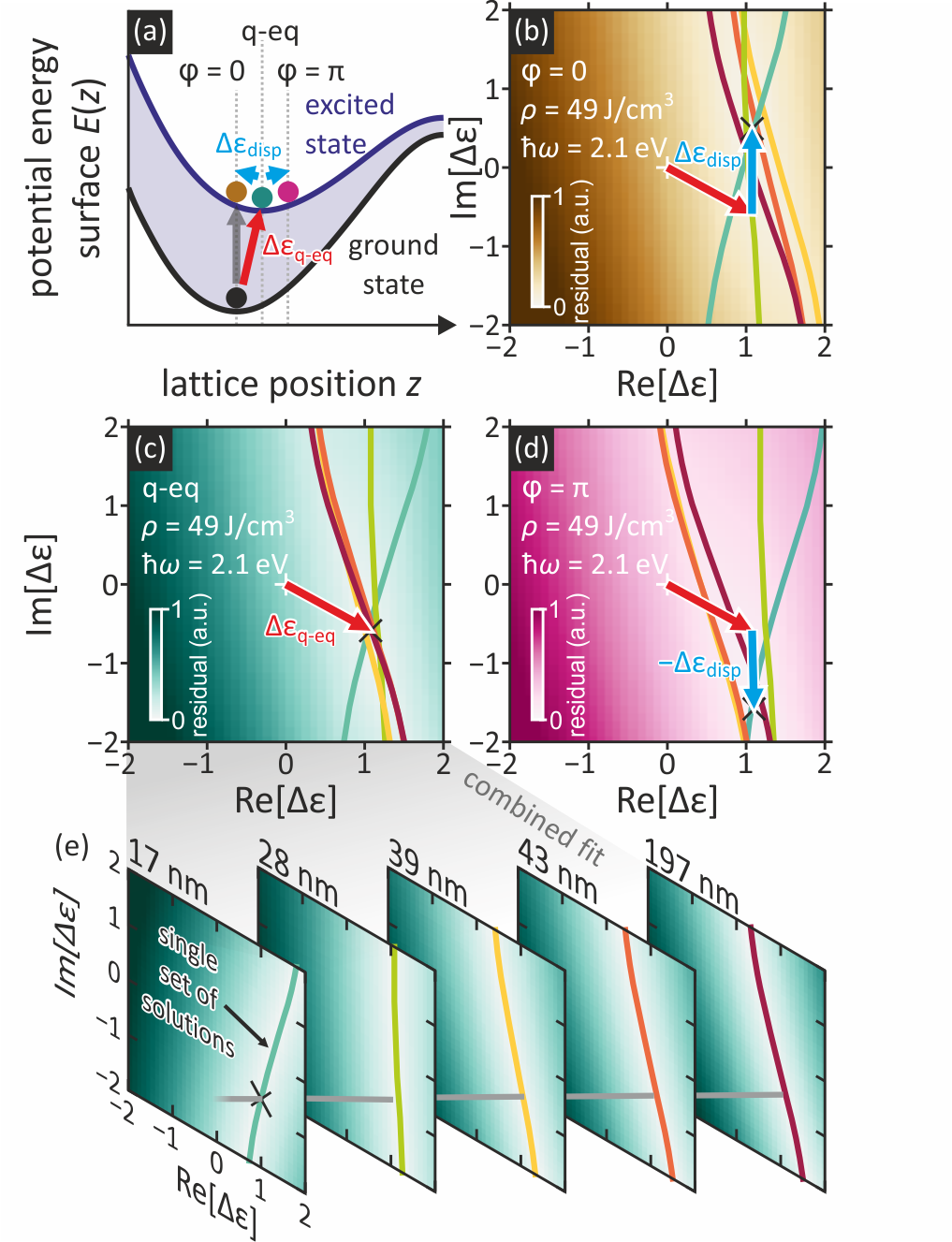}
		\caption{
			(Color online) (a) Bismuth’s potential energy surface according to \cite{Murray2005}. The colored dots refer to the positions of the oscillation shown in Fig. \ref{fig:one} (b). The grey arrow marks the pathway of excitation. The red and blue arrows indicate the changes to the quasi-equilibrium position (q-eq) and the displacive component (disp) respectively. (b)-(d) combined residual for all films at the positions marked in (a).  (e) residuals of the fits for $\Delta \varepsilon (d)$ at the q-eq position.
		}
		\label{fig:two}
	\end{figure}
	
	Owing to the cosine approximation for $\Delta R/R_0(\Delta t)$, it is sufficient to recover the change of dielectric function $\Delta \varepsilon$ at the two extrema and the node of the oscillations. Thus $A_\mathrm{bg} + A_\mathrm{osc}$, $A_\mathrm{bg}$ and $A_\mathrm{bg} - A_\mathrm{osc}$ refer to the dielectric function changes $\Delta \varepsilon_0$, $\Delta \varepsilon_\mathrm{q-eq}$ and $\Delta \varepsilon_\pi$ at the three indicated positions. q-eq denotes the new quasi equilibrium position and the phases $\varphi=0$ and $\varphi=\pi$ describe the maxima and minima of the oscillation. The three distinct points are marked in the potential energy surface (described in Ref. \cite{Murray2005}) by colored dots (see Fig. \ref{fig:two}).
	
	The dielectric function change $\Delta \varepsilon$ is obtained by fitting $\Delta R/R_0 (\Delta \varepsilon, d, \hbar\omega)$ at the three distinct points in the potential energy surface. Because $\Delta R/R_0$ is connected to $\Delta \varepsilon$ by the absolute square and all complex information is lost, an unambiguous solution for $\Delta \varepsilon$ is only obtained by providing additional information. This can be transmission \cite{Obergfell2020}, polarisation \cite{Boschini2015, Richter2021}, incident angle \cite{Roeser2003, Richter2021} or in this case: the film thickness $d$. The use of $d$ is only reasonable when thin film interference causes significant changes of $\Delta R/R_0$ within a regime of $d$ where the films are homogeneously excited. We can exclude any optical property changes of the Si substrate as a Schottky barrier at the Bi/Si interface \cite{Hricovini1992} prevents its excitation. The formation of the Bi surface state is prevented by the formation of a $3\,\mathrm{nm}$ thick oxide layer{\cite{Payer2012}}. Quantum confinement effects are negligible for the film thicknesses studied herein {\cite{Renzi1993}}. Therefore, $\Delta R/R_0(\Delta\varepsilon, d, \hbar \omega)$ is computed solely with the Fresnel coefficient at the air/Bi and Bi/Si interfaces, $d$ and the dielectric function change in the bismuth film $\Delta \varepsilon$, using the transfer matrix method \cite{Katsidis2002}.
	
	The individual panels in Fig. \ref{fig:two} (e) show exemplarily the solutions, indicated by the fits' residuals $|A_\mathrm{bg} (d) - \Delta R/R_0(d, \Delta \varepsilon)|$, for the change in the real and imaginary parts of the dielectric function for five different film thicknesses at the same $\hbar \omega=2.1\,\mathrm{eV}$. No unambiguous solution for $\Delta \varepsilon = \mathrm{Re}\!\left[\Delta \varepsilon\right] + i\,\mathrm{Im}\!\left[\Delta \varepsilon\right]$ is obtained for each of the film thicknesses $d$, but a set of solutions in the $\mathrm{Re}\!\left[\Delta \varepsilon\right]$-$\mathrm{Im}\!\left[\Delta \varepsilon\right]$-plane (indicated by a solid line in Fig. \ref{fig:two} (e)) that changes due to thin film interference. Based on the safe assumption that at the same absorbed energy density, $\Delta \varepsilon$ is independent on film thickness $d$, the intersection of these individual solutions (grey line) yields the unambiguous solution for $\Delta \varepsilon$.
	
	Figure \ref{fig:two} (b)-(d) summarize this procedure for $\Delta \varepsilon_0$, $\Delta \varepsilon_\mathrm{q-eq}$ and $\Delta \varepsilon_\pi$. Correspondingly to panel (e), the colored lines mark the individual solutions for each film thickness $d$ while the cross identifies the overall minimum of the combined residuals $\sum_d |A (d) - \Delta R/R_0(d, \Delta \varepsilon)|$. Across Fig. \ref{fig:two} $\Delta \varepsilon_\mathrm{q-eq}$ is emphasized by red arrows. It is useful to define a displacive component $\Delta \varepsilon_\mathrm{disp}$ (blue arrows in Fig. \ref{fig:two}), including only the changes from q-eq towards the maximum ($\varphi=0$) and minimum ($\varphi=\pi$) displacement within the potential energy surface. $\Delta \varepsilon_\mathrm{disp}$ is modulated solely by the nuclear motion in an excited state. Because of the cosine approximation $\Delta \varepsilon_\mathrm{disp}$ is symmetric:
	\begin{equation}
		\Delta \varepsilon_\mathrm{disp} = \Delta \varepsilon_0 - \Delta \varepsilon_\mathrm{q-eq}=-\Delta \varepsilon_\pi + \Delta \varepsilon_\mathrm{q-eq}\,.
	\end{equation}
	
	
	\begin{figure}[t!]
		\centering\includegraphics[width=1.0\columnwidth]{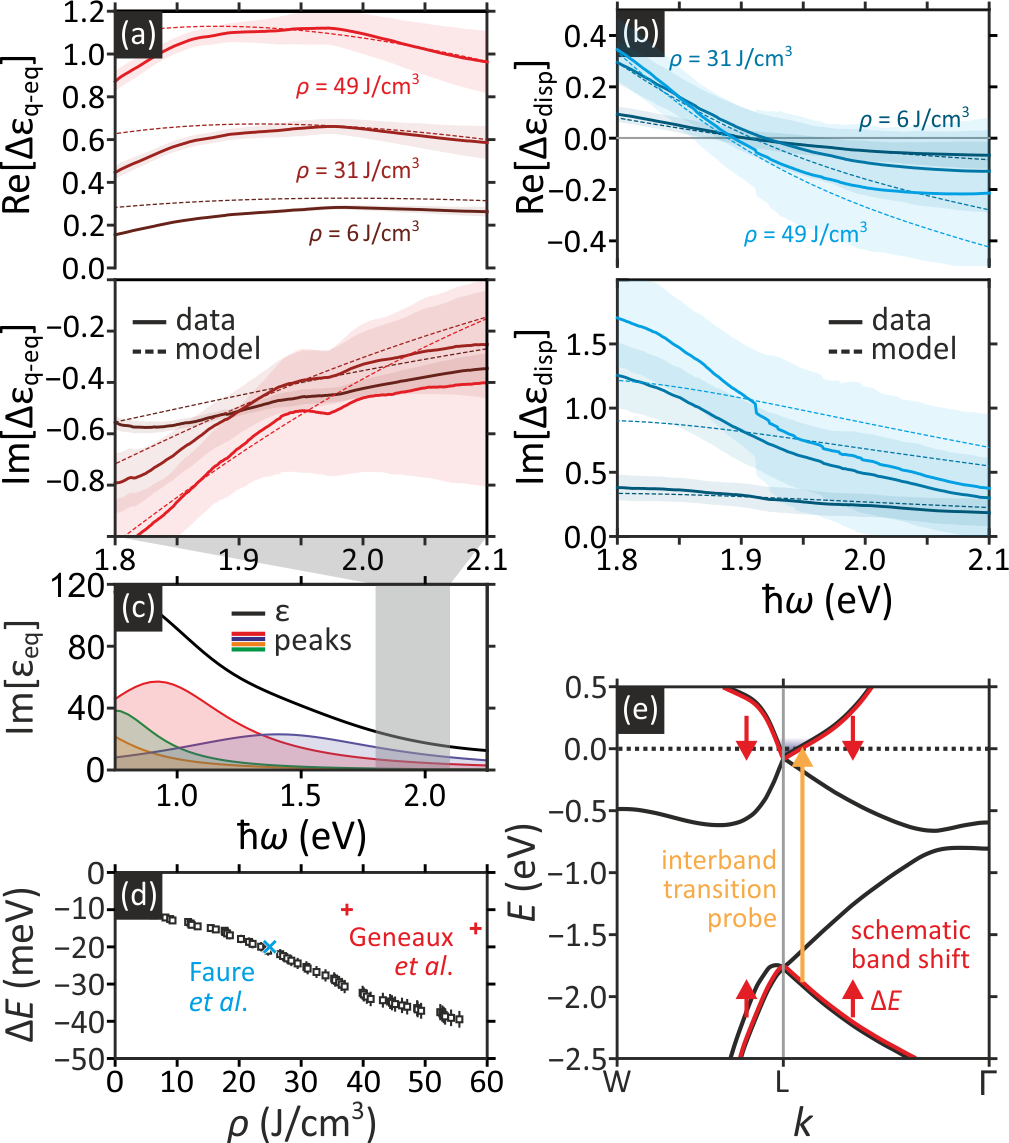}
		\caption{
			(Color online) (a) change of real and imaginary part of $\Delta \varepsilon_\mathrm{q-eq}$ for selected absorbed energy densities $\rho$. Solid lines mark $\Delta \varepsilon$, the light-colored area the uncertainty and the dashed lines the fitted model. (b) $\Delta \varepsilon_\mathrm{disp}$ for selected $\rho$. (c) Imaginary part of bismuth’s dielectric function $\varepsilon_\mathrm{eq}$ (black), based on the model from Ref. \cite{Toudert2017}. The colored peaks are the single Lorentz peaks. (d) Peak shift fit parameter $\Delta E$. (e) sketch of bismuth’s band structure based on Ref. \cite{Aguilera2015}. The orange arrow marks a possible probe interband transition. The blue area indicates population changes affecting the probe intensity. The shift observed in $\Delta E$ is caused by arbitrary band shifts, indicated in red.
		}\label{fig:three}
	\end{figure}
	
	This procedure to obtain $\Delta \varepsilon$ was performed for all film thicknesses $d$, photon energies $\hbar\omega$, and absorbed energy densities $\rho$ at the positions \emph{q-eq} and \emph{disp}, yielding an indigestible multi dimensional parameter space. Displaying only selected energy densities $\rho$ provides more insight. Figure \ref{fig:three} (a) shows $\Delta \varepsilon_\mathrm{q-eq} ( \hbar\omega)$ for selected $\rho$ and Fig. \ref{fig:three} (b) $\Delta \varepsilon_\mathrm{disp} (\hbar\omega)$ respectively. The solid lines in Fig. 3 (a) and (b) depict the real and imaginary part of $\Delta \varepsilon (\hbar\omega)$. The light-colored areas mark the regime of uncertainty obtained from the fits. Figure. \ref{fig:three} (c) shows the imaginary part of the equilibrium dielectric function $\varepsilon_\mathrm{eq}$ obtained over a broader spectral range by ellipsometry measurements based on a model taken from literature \cite{Toudert2017}.
	
	The next step is to provide a model for the observed changes in the dielectric function $\Delta \varepsilon_\mathrm{q-eq}$ and $\Delta \varepsilon_\mathrm{disp}$, that gives insight into the transient electronic dynamics in the material. 
	Since the overall excitation densities $\rho$ are weak and hence deviations from equilibrium are small, we construct a simple model by changing the dielectric function under equilibrium conditions 
    \begin{equation}
        \varepsilon(\rho) = \varepsilon_\mathrm{eq} + \Delta\varepsilon_\mathrm{q-eq}(\rho) + \Delta\varepsilon_\mathrm{disp}(\rho).
    \end{equation}
	Drude or intraband contributions (dominant in metals) are negligible for Bi in the range of visible light \cite{Timrov2012} and thus $\varepsilon_\mathrm{eq}$ is mainly defined by interband transitions {\cite{Cardona1964, Hunderi1975, Toudert2017}} In such cases the Lorentz oscillator model \cite{Liu2020, Han2023} is often used, and defined as the sum of single complex Lorentz peaks that obey the Kramers-Kronig relation
	\begin{equation}
		\varepsilon_\mathrm{eq}(\hbar\omega)=\sum_j \frac{A_j \Gamma_j, E_j}{E_j^2 - \hbar\omega^2-i\Gamma_j\hbar\omega}\,,
	\end{equation}
	with the parameters amplitude $A_j$, width $\Gamma_j$, and central energy $E_j$. The numerical values were taken from literature \cite{Toudert2017} and agree with our ellipsometry data (see Supplemental Material). Comparing the Lorentz peaks with the spectral window studied (see Fig. {\ref{fig:three}} (c)), we notice that only the tails of two oscillators are important. In a simple picture the oscillators refer to interband transitions with large spectral weight in the band structure.
	
	First, we consider only the influence of the excited charge carriers under a slightly weakened Peierls distortion of the lattice at the new quasi-equilibrium position (q-eq), treating the system as if there were no coherent phonons. The modification of Lorentz oscillators has proven to describe the impact of population changes and band shifts in the optical properties, as recently demonstrated in transition metal dichalcogenides \cite{Calati2021, Smejkal2022, Trovatello2022}. In most materials the parameters of the Lorentz oscillators do not vary much throughout a broad spectral range \cite{Vina1984, Lautenschlager1987, Lautenschlager1987a, Shkrebtii2010} and since our spectral range is comparatively small, we have to approximate the change of all Lorentz peaks with three parameters only: $A_j\rightarrow f_A A_j$, $\Gamma_j\rightarrow f_\Gamma \Gamma_j$  and $E_j\rightarrow E_j + \Delta E$. They are determined by fitting the modified Lorentz oscillators (dashed lines) to the data shown in Fig. \ref{fig:three} (a) for each corresponding $\rho$. The change in $\Delta E$ as a function of energy density is shown in Fig. \ref{fig:three} (d). We find that the Lorentz peaks decrease in amplitude, broaden and redshift in energy. This result is very close to the general behavior of the parameters $A_j$, $\Gamma_j$ and $E_j$, describing the change in dielectric function in semiconductors upon changes in temperature as observed for Si, Ge, GaAs \cite{Vina1984, Lautenschlager1987, Lautenschlager1987a, Shkrebtii2010}.
    
    In these references however, the carriers are thermally and not optically excited and band structure changes arise from thermal lattice expansion instead of a modified Peierls-like distortion. This is still comparable to our situation, because in Bi optically excited electrons are quasi-thermalized 150\,fs after excitation \cite{Faure2013}, matching the time window studied (see Fig. \ref{fig:one} (b)). Also, the lattice changes induced by thermal heating are related to a reduction of the Peierls distortion and softening of the energy surface, respectively \cite{Hase1998}. Furthermore, we can compare our results to energy shifts $\Delta E$ observed in time-resolved angle-resolved photoemission spectroscopy \cite{Faure2013} and UV-absorption spectroscopy studies\cite{Geneaux2021}. Both, compared to the fitted $\Delta E$, exhibit the same trend and the same order of magnitude, while absolute values are hard to compare due to the ambiguity of excited carrier density \cite{Jnawali2021, Thiemann2022}.
	
	The relation between the behavior of excited carriers, $\Delta E$ and modification of peak amplitudes and widths can be explained in the band structure picture as sketched around the \emph{L}-point in Fig. \ref{fig:three} (e). Population dynamics upon photoexcitation lead to bleaching and thus to changes of the maximum in $\mathrm{Re}[\Delta \varepsilon_\mathrm{q-eq}]$ around 2\,eV. During thermalization the carriers relax to the band edge and accumulate at the \emph{L}-point (electrons) and \emph{T}-point (holes) after 150\,fs respectively \cite{Faure2013}. This region is exactly covered by out spectral range. The $L_a(2)\rightarrow L_s(3)$ transition (orange arrow in Fig. \ref{fig:three} (e)) connects an occupied valence band to this partially occupied conduction band minimum, $\Delta E \approx$ 2\,eV apart \cite{Hunderi1975, Liu1995, Aguilera2015, Koenig2021}. With this transition bleached by excited carriers, thus reducing the availability of final states, the absorption decreases the strongest at $\approx$2\,eV. Furthermore, the increased peak width can be understood in a two-temperature model through the smeared-out Fermi-distribution of the thermalized but still hot carriers \cite{Anisimov1974}.

	Moving on to the displacive component, $\Delta \varepsilon_\mathrm{disp}$. Based on previous studies on Sb by Stevens \emph{et al.} \cite{Stevens2002}, which treated the coherent phonon excitations without excitation of the electron system, we would expect a Raman correction term. This term, $\Delta \varepsilon_\mathrm{disp}=\chi_R Q(t=2\pi/\Omega)$, is described by the Raman susceptibility $\chi_R$ and the phonon coordinate $Q(t)$. The Raman susceptibility, $\chi_R$, can be described in terms of a second-order nonlinear interaction between the probe light and a "phonon field", which is coupled to the electron system via the electron-phonon coupling, $\Xi$. Using the simplifications $\Omega\ll\omega$ and $\Xi=\mathrm{const}$ \cite{Cardona1975, Stevens2002}, we obtain
    \begin{equation}
        \Delta \varepsilon_\mathrm{disp} = \chi_R Q \approx \frac{\Xi Q}{4\pi\hbar} \frac{\mathrm{d}}{\mathrm{d}\omega} \varepsilon_\mathrm{eq}.
    \end{equation}
    In contrast, we find the non-empirical relation
    \begin{equation}
        \Delta \varepsilon_\mathrm{disp}\propto \frac{\mathrm{d}}{\mathrm{d}\omega} \Delta\varepsilon_\mathrm{q-eq},
    \end{equation}
    which seems to depend on the quasi-equilibrium component or the change of $\varepsilon_\mathrm{eq}$ in general. An explanation might be that population changes at the \emph{L}-point in Sb is do not strongly affect interband transitions around 2\,eV, as in bismuth. Additionally, the assumption $\Xi=\mathrm{const}$ may not necessarily hold, and the previous simplification does not account for population changes. Therefore, we hypothesize that the population of excited carriers play a significant, yet not fully identified, role in the description of optical properties changes by coherent phonons in Bi. Proposing a first principle solution to this open question based on the available data set would be highly speculative.

	In summary, we tracked the changes in the complex dielectric function $\Delta \varepsilon$ of photoexcited Bi and separated it into a quasi-equilibrium and a displacive structural component, incorporating the carrier and lattice dynamics. These contributions were analyzed using a phenomenological approach. A modified Lorentz oscillator model reveals a substantial impact of the interband transition to the conduction band minimum at the $L$ point, and the effects of the electron gas temperature and Pauli blocking on the optical properties of this strongly correlated system.

	\begin{acknowledgements}
		We gratefully acknowledge fruitful discussions with R. Merlin, M. Henstridge, and A. von Hoegen. Funding by the Deutsche Forschungsgemeinschaft (DFG, German Research Foundation) through project B04 and C03 of Collaborative Research Center SFB 1242 “Nonequilibrium dynamics of condensed matter in the time domain” (Project\nobreakdash-ID~278162697) is appreciated.
		G.S. acknowledges the support of the National Science and Engineering Research Council of Canada, the Canada Foundation for Innovation and Ontario Research Fund. The authors declare no competing financial interest.
	\end{acknowledgements}

	\bibliography{main} 

\begin{thebibliography}{57}%
\makeatletter
\providecommand \@ifxundefined [1]{%
 \@ifx{#1\undefined}
}%
\providecommand \@ifnum [1]{%
 \ifnum #1\expandafter \@firstoftwo
 \else \expandafter \@secondoftwo
 \fi
}%
\providecommand \@ifx [1]{%
 \ifx #1\expandafter \@firstoftwo
 \else \expandafter \@secondoftwo
 \fi
}%
\providecommand \natexlab [1]{#1}%
\providecommand \enquote  [1]{``#1''}%
\providecommand \bibnamefont  [1]{#1}%
\providecommand \bibfnamefont [1]{#1}%
\providecommand \citenamefont [1]{#1}%
\providecommand \href@noop [0]{\@secondoftwo}%
\providecommand \href [0]{\begingroup \@sanitize@url \@href}%
\providecommand \@href[1]{\@@startlink{#1}\@@href}%
\providecommand \@@href[1]{\endgroup#1\@@endlink}%
\providecommand \@sanitize@url [0]{\catcode `\\12\catcode `\$12\catcode
  `\&12\catcode `\#12\catcode `\^12\catcode `\_12\catcode `\%12\relax}%
\providecommand \@@startlink[1]{}%
\providecommand \@@endlink[0]{}%
\providecommand \url  [0]{\begingroup\@sanitize@url \@url }%
\providecommand \@url [1]{\endgroup\@href {#1}{\urlprefix }}%
\providecommand \urlprefix  [0]{URL }%
\providecommand \Eprint [0]{\href }%
\providecommand \doibase [0]{https://doi.org/}%
\providecommand \selectlanguage [0]{\@gobble}%
\providecommand \bibinfo  [0]{\@secondoftwo}%
\providecommand \bibfield  [0]{\@secondoftwo}%
\providecommand \translation [1]{[#1]}%
\providecommand \BibitemOpen [0]{}%
\providecommand \bibitemStop [0]{}%
\providecommand \bibitemNoStop [0]{.\EOS\space}%
\providecommand \EOS [0]{\spacefactor3000\relax}%
\providecommand \BibitemShut  [1]{\csname bibitem#1\endcsname}%
\let\auto@bib@innerbib\@empty
\bibitem [{\citenamefont {Schoenlein}\ \emph {et~al.}(1987)\citenamefont
  {Schoenlein}, \citenamefont {Lin}, \citenamefont {Fujimoto},\ and\
  \citenamefont {Eesley}}]{Schoenlein1987}%
  \BibitemOpen
  \bibfield  {author} {\bibinfo {author} {\bibfnamefont {R.~W.}\ \bibnamefont
  {Schoenlein}}, \bibinfo {author} {\bibfnamefont {W.~Z.}\ \bibnamefont {Lin}},
  \bibinfo {author} {\bibfnamefont {J.~G.}\ \bibnamefont {Fujimoto}},\ and\
  \bibinfo {author} {\bibfnamefont {G.~L.}\ \bibnamefont {Eesley}},\ }\href
  {https://doi.org/10.1103/physrevlett.58.1680} {\bibfield  {journal} {\bibinfo
   {journal} {Physical Review Letters}\ }\textbf {\bibinfo {volume} {58}},\
  \bibinfo {pages} {1680} (\bibinfo {year} {1987})}\BibitemShut {NoStop}%
\bibitem [{\citenamefont {Cheng}\ \emph {et~al.}(1990)\citenamefont {Cheng},
  \citenamefont {Brorson}, \citenamefont {Kazeroonian}, \citenamefont
  {Moodera}, \citenamefont {Dresselhaus}, \citenamefont {Dresselhaus},\ and\
  \citenamefont {Ippen}}]{Cheng1990}%
  \BibitemOpen
  \bibfield  {author} {\bibinfo {author} {\bibfnamefont {T.~K.}\ \bibnamefont
  {Cheng}}, \bibinfo {author} {\bibfnamefont {S.~D.}\ \bibnamefont {Brorson}},
  \bibinfo {author} {\bibfnamefont {A.~S.}\ \bibnamefont {Kazeroonian}},
  \bibinfo {author} {\bibfnamefont {J.~S.}\ \bibnamefont {Moodera}}, \bibinfo
  {author} {\bibfnamefont {G.}~\bibnamefont {Dresselhaus}}, \bibinfo {author}
  {\bibfnamefont {M.~S.}\ \bibnamefont {Dresselhaus}},\ and\ \bibinfo {author}
  {\bibfnamefont {E.~P.}\ \bibnamefont {Ippen}},\ }\href
  {https://doi.org/10.1063/1.104090} {\bibfield  {journal} {\bibinfo  {journal}
  {Applied Physics Letters}\ }\textbf {\bibinfo {volume} {57}},\ \bibinfo
  {pages} {1004} (\bibinfo {year} {1990})}\BibitemShut {NoStop}%
\bibitem [{\citenamefont {Obergfell}\ and\ \citenamefont
  {Demsar}(2020)}]{Obergfell2020}%
  \BibitemOpen
  \bibfield  {author} {\bibinfo {author} {\bibfnamefont {M.}~\bibnamefont
  {Obergfell}}\ and\ \bibinfo {author} {\bibfnamefont {J.}~\bibnamefont
  {Demsar}},\ }\href {https://doi.org/10.1103/physrevlett.124.037401}
  {\bibfield  {journal} {\bibinfo  {journal} {Physical Review Letters}\
  }\textbf {\bibinfo {volume} {124}},\ \bibinfo {pages} {037401} (\bibinfo
  {year} {2020})}\BibitemShut {NoStop}%
\bibitem [{\citenamefont {Calati}\ \emph {et~al.}(2021)\citenamefont {Calati},
  \citenamefont {Li}, \citenamefont {Zhu},\ and\ \citenamefont
  {Stähler}}]{Calati2021}%
  \BibitemOpen
  \bibfield  {author} {\bibinfo {author} {\bibfnamefont {S.}~\bibnamefont
  {Calati}}, \bibinfo {author} {\bibfnamefont {Q.}~\bibnamefont {Li}}, \bibinfo
  {author} {\bibfnamefont {X.}~\bibnamefont {Zhu}},\ and\ \bibinfo {author}
  {\bibfnamefont {J.}~\bibnamefont {Stähler}},\ }\href
  {https://doi.org/10.1039/d1cp03437e} {\bibfield  {journal} {\bibinfo
  {journal} {Physical Chemistry Chemical Physics}\ }\textbf {\bibinfo {volume}
  {23}},\ \bibinfo {pages} {22640} (\bibinfo {year} {2021})}\BibitemShut
  {NoStop}%
\bibitem [{\citenamefont {Gerber}\ \emph {et~al.}(2017)\citenamefont {Gerber},
  \citenamefont {Yang}, \citenamefont {Zhu}, \citenamefont {Soifer},
  \citenamefont {Sobota}, \citenamefont {Rebec}, \citenamefont {Lee},
  \citenamefont {Jia}, \citenamefont {Moritz}, \citenamefont {Jia},
  \citenamefont {Gauthier}, \citenamefont {Li}, \citenamefont {Leuenberger},
  \citenamefont {Zhang}, \citenamefont {Chaix}, \citenamefont {Li},
  \citenamefont {Jang}, \citenamefont {Lee}, \citenamefont {Yi}, \citenamefont
  {Dakovski}, \citenamefont {Song}, \citenamefont {Glownia}, \citenamefont
  {Nelson}, \citenamefont {Kim}, \citenamefont {Chuang}, \citenamefont
  {Hussain}, \citenamefont {Moore}, \citenamefont {Devereaux}, \citenamefont
  {Lee}, \citenamefont {Kirchmann},\ and\ \citenamefont {Shen}}]{Gerber2017}%
  \BibitemOpen
  \bibfield  {author} {\bibinfo {author} {\bibfnamefont {S.}~\bibnamefont
  {Gerber}}, \bibinfo {author} {\bibfnamefont {S.-L.}\ \bibnamefont {Yang}},
  \bibinfo {author} {\bibfnamefont {D.}~\bibnamefont {Zhu}}, \bibinfo {author}
  {\bibfnamefont {H.}~\bibnamefont {Soifer}}, \bibinfo {author} {\bibfnamefont
  {J.~A.}\ \bibnamefont {Sobota}}, \bibinfo {author} {\bibfnamefont
  {S.}~\bibnamefont {Rebec}}, \bibinfo {author} {\bibfnamefont {J.~J.}\
  \bibnamefont {Lee}}, \bibinfo {author} {\bibfnamefont {T.}~\bibnamefont
  {Jia}}, \bibinfo {author} {\bibfnamefont {B.}~\bibnamefont {Moritz}},
  \bibinfo {author} {\bibfnamefont {C.}~\bibnamefont {Jia}}, \bibinfo {author}
  {\bibfnamefont {A.}~\bibnamefont {Gauthier}}, \bibinfo {author}
  {\bibfnamefont {Y.}~\bibnamefont {Li}}, \bibinfo {author} {\bibfnamefont
  {D.}~\bibnamefont {Leuenberger}}, \bibinfo {author} {\bibfnamefont
  {Y.}~\bibnamefont {Zhang}}, \bibinfo {author} {\bibfnamefont
  {L.}~\bibnamefont {Chaix}}, \bibinfo {author} {\bibfnamefont
  {W.}~\bibnamefont {Li}}, \bibinfo {author} {\bibfnamefont {H.}~\bibnamefont
  {Jang}}, \bibinfo {author} {\bibfnamefont {J.-S.}\ \bibnamefont {Lee}},
  \bibinfo {author} {\bibfnamefont {M.}~\bibnamefont {Yi}}, \bibinfo {author}
  {\bibfnamefont {G.~L.}\ \bibnamefont {Dakovski}}, \bibinfo {author}
  {\bibfnamefont {S.}~\bibnamefont {Song}}, \bibinfo {author} {\bibfnamefont
  {J.~M.}\ \bibnamefont {Glownia}}, \bibinfo {author} {\bibfnamefont
  {S.}~\bibnamefont {Nelson}}, \bibinfo {author} {\bibfnamefont {K.~W.}\
  \bibnamefont {Kim}}, \bibinfo {author} {\bibfnamefont {Y.-D.}\ \bibnamefont
  {Chuang}}, \bibinfo {author} {\bibfnamefont {Z.}~\bibnamefont {Hussain}},
  \bibinfo {author} {\bibfnamefont {R.~G.}\ \bibnamefont {Moore}}, \bibinfo
  {author} {\bibfnamefont {T.~P.}\ \bibnamefont {Devereaux}}, \bibinfo {author}
  {\bibfnamefont {W.-S.}\ \bibnamefont {Lee}}, \bibinfo {author} {\bibfnamefont
  {P.~S.}\ \bibnamefont {Kirchmann}},\ and\ \bibinfo {author} {\bibfnamefont
  {Z.-X.}\ \bibnamefont {Shen}},\ }\href
  {https://doi.org/10.1126/science.aak9946} {\bibfield  {journal} {\bibinfo
  {journal} {Science}\ }\textbf {\bibinfo {volume} {357}},\ \bibinfo {pages}
  {71} (\bibinfo {year} {2017})}\BibitemShut {NoStop}%
\bibitem [{\citenamefont {Otto}\ \emph {et~al.}(2018)\citenamefont {Otto},
  \citenamefont {de~Cotret}, \citenamefont {Valverde-Chavez}, \citenamefont
  {Tiwari}, \citenamefont {{\'{E}}mond}, \citenamefont {Chaker}, \citenamefont
  {Cooke},\ and\ \citenamefont {Siwick}}]{Otto2018}%
  \BibitemOpen
  \bibfield  {author} {\bibinfo {author} {\bibfnamefont {M.~R.}\ \bibnamefont
  {Otto}}, \bibinfo {author} {\bibfnamefont {L.~P.~R.}\ \bibnamefont
  {de~Cotret}}, \bibinfo {author} {\bibfnamefont {D.~A.}\ \bibnamefont
  {Valverde-Chavez}}, \bibinfo {author} {\bibfnamefont {K.~L.}\ \bibnamefont
  {Tiwari}}, \bibinfo {author} {\bibfnamefont {N.}~\bibnamefont {{\'{E}}mond}},
  \bibinfo {author} {\bibfnamefont {M.}~\bibnamefont {Chaker}}, \bibinfo
  {author} {\bibfnamefont {D.~G.}\ \bibnamefont {Cooke}},\ and\ \bibinfo
  {author} {\bibfnamefont {B.~J.}\ \bibnamefont {Siwick}},\ }\href
  {https://doi.org/10.1073/pnas.1808414115} {\bibfield  {journal} {\bibinfo
  {journal} {Proceedings of the National Academy of Sciences}\ }\textbf
  {\bibinfo {volume} {116}},\ \bibinfo {pages} {450} (\bibinfo {year}
  {2018})}\BibitemShut {NoStop}%
\bibitem [{\citenamefont {Demsar}\ \emph {et~al.}(2002)\citenamefont {Demsar},
  \citenamefont {Forr{\'{o}}}, \citenamefont {Berger},\ and\ \citenamefont
  {Mihailovic}}]{Demsar2002}%
  \BibitemOpen
  \bibfield  {author} {\bibinfo {author} {\bibfnamefont {J.}~\bibnamefont
  {Demsar}}, \bibinfo {author} {\bibfnamefont {L.}~\bibnamefont {Forr{\'{o}}}},
  \bibinfo {author} {\bibfnamefont {H.}~\bibnamefont {Berger}},\ and\ \bibinfo
  {author} {\bibfnamefont {D.}~\bibnamefont {Mihailovic}},\ }\href
  {https://doi.org/10.1103/physrevb.66.041101} {\bibfield  {journal} {\bibinfo
  {journal} {Physical Review B}\ }\textbf {\bibinfo {volume} {66}},\ \bibinfo
  {pages} {041101} (\bibinfo {year} {2002})}\BibitemShut {NoStop}%
\bibitem [{\citenamefont {Wall}\ \emph {et~al.}(2012)\citenamefont {Wall},
  \citenamefont {Wegkamp}, \citenamefont {Foglia}, \citenamefont {Appavoo},
  \citenamefont {Nag}, \citenamefont {Haglund}, \citenamefont {Stähler},\ and\
  \citenamefont {Wolf}}]{Wall2012}%
  \BibitemOpen
  \bibfield  {author} {\bibinfo {author} {\bibfnamefont {S.}~\bibnamefont
  {Wall}}, \bibinfo {author} {\bibfnamefont {D.}~\bibnamefont {Wegkamp}},
  \bibinfo {author} {\bibfnamefont {L.}~\bibnamefont {Foglia}}, \bibinfo
  {author} {\bibfnamefont {K.}~\bibnamefont {Appavoo}}, \bibinfo {author}
  {\bibfnamefont {J.}~\bibnamefont {Nag}}, \bibinfo {author} {\bibfnamefont
  {R.}~\bibnamefont {Haglund}}, \bibinfo {author} {\bibfnamefont
  {J.}~\bibnamefont {Stähler}},\ and\ \bibinfo {author} {\bibfnamefont
  {M.}~\bibnamefont {Wolf}},\ }\bibfield  {journal} {\bibinfo  {journal}
  {Nature Communications}\ }\textbf {\bibinfo {volume} {3}},\ \href
  {https://doi.org/10.1038/ncomms1719} {10.1038/ncomms1719} (\bibinfo {year}
  {2012})\BibitemShut {NoStop}%
\bibitem [{\citenamefont {Cheng}\ \emph {et~al.}(1991)\citenamefont {Cheng},
  \citenamefont {Vidal}, \citenamefont {Zeiger}, \citenamefont {Dresselhaus},
  \citenamefont {Dresselhaus},\ and\ \citenamefont {Ippen}}]{Cheng1991}%
  \BibitemOpen
  \bibfield  {author} {\bibinfo {author} {\bibfnamefont {T.~K.}\ \bibnamefont
  {Cheng}}, \bibinfo {author} {\bibfnamefont {J.}~\bibnamefont {Vidal}},
  \bibinfo {author} {\bibfnamefont {H.~J.}\ \bibnamefont {Zeiger}}, \bibinfo
  {author} {\bibfnamefont {G.}~\bibnamefont {Dresselhaus}}, \bibinfo {author}
  {\bibfnamefont {M.~S.}\ \bibnamefont {Dresselhaus}},\ and\ \bibinfo {author}
  {\bibfnamefont {E.~P.}\ \bibnamefont {Ippen}},\ }\href
  {https://doi.org/10.1063/1.106187} {\bibfield  {journal} {\bibinfo  {journal}
  {Applied Physics Letters}\ }\textbf {\bibinfo {volume} {59}},\ \bibinfo
  {pages} {1923} (\bibinfo {year} {1991})}\BibitemShut {NoStop}%
\bibitem [{\citenamefont {Hase}\ \emph {et~al.}(1996)\citenamefont {Hase},
  \citenamefont {Mizoguchi}, \citenamefont {Harima}, \citenamefont {Nakashima},
  \citenamefont {Tani}, \citenamefont {Sakai},\ and\ \citenamefont
  {Hangyo}}]{Hase1996}%
  \BibitemOpen
  \bibfield  {author} {\bibinfo {author} {\bibfnamefont {M.}~\bibnamefont
  {Hase}}, \bibinfo {author} {\bibfnamefont {K.}~\bibnamefont {Mizoguchi}},
  \bibinfo {author} {\bibfnamefont {H.}~\bibnamefont {Harima}}, \bibinfo
  {author} {\bibfnamefont {S.}~\bibnamefont {Nakashima}}, \bibinfo {author}
  {\bibfnamefont {M.}~\bibnamefont {Tani}}, \bibinfo {author} {\bibfnamefont
  {K.}~\bibnamefont {Sakai}},\ and\ \bibinfo {author} {\bibfnamefont
  {M.}~\bibnamefont {Hangyo}},\ }\href {https://doi.org/10.1063/1.117502}
  {\bibfield  {journal} {\bibinfo  {journal} {Applied Physics Letters}\
  }\textbf {\bibinfo {volume} {69}},\ \bibinfo {pages} {2474} (\bibinfo {year}
  {1996})}\BibitemShut {NoStop}%
\bibitem [{\citenamefont {DeCamp}\ \emph {et~al.}(2001)\citenamefont {DeCamp},
  \citenamefont {Reis}, \citenamefont {Bucksbaum},\ and\ \citenamefont
  {Merlin}}]{DeCamp2001}%
  \BibitemOpen
  \bibfield  {author} {\bibinfo {author} {\bibfnamefont {M.~F.}\ \bibnamefont
  {DeCamp}}, \bibinfo {author} {\bibfnamefont {D.~A.}\ \bibnamefont {Reis}},
  \bibinfo {author} {\bibfnamefont {P.~H.}\ \bibnamefont {Bucksbaum}},\ and\
  \bibinfo {author} {\bibfnamefont {R.}~\bibnamefont {Merlin}},\ }\href
  {https://doi.org/10.1103/physrevb.64.092301} {\bibfield  {journal} {\bibinfo
  {journal} {Physical Review B}\ }\textbf {\bibinfo {volume} {64}},\ \bibinfo
  {pages} {092301} (\bibinfo {year} {2001})}\BibitemShut {NoStop}%
\bibitem [{\citenamefont {Boschetto}\ \emph {et~al.}(2008)\citenamefont
  {Boschetto}, \citenamefont {Gamaly}, \citenamefont {Rode}, \citenamefont
  {Luther-Davies}, \citenamefont {Glijer}, \citenamefont {Garl}, \citenamefont
  {Albert}, \citenamefont {Rousse},\ and\ \citenamefont
  {Etchepare}}]{Boschetto2008}%
  \BibitemOpen
  \bibfield  {author} {\bibinfo {author} {\bibfnamefont {D.}~\bibnamefont
  {Boschetto}}, \bibinfo {author} {\bibfnamefont {E.~G.}\ \bibnamefont
  {Gamaly}}, \bibinfo {author} {\bibfnamefont {A.~V.}\ \bibnamefont {Rode}},
  \bibinfo {author} {\bibfnamefont {B.}~\bibnamefont {Luther-Davies}}, \bibinfo
  {author} {\bibfnamefont {D.}~\bibnamefont {Glijer}}, \bibinfo {author}
  {\bibfnamefont {T.}~\bibnamefont {Garl}}, \bibinfo {author} {\bibfnamefont
  {O.}~\bibnamefont {Albert}}, \bibinfo {author} {\bibfnamefont
  {A.}~\bibnamefont {Rousse}},\ and\ \bibinfo {author} {\bibfnamefont
  {J.}~\bibnamefont {Etchepare}},\ }\href
  {https://doi.org/10.1103/physrevlett.100.027404} {\bibfield  {journal}
  {\bibinfo  {journal} {Physical Review Letters}\ }\textbf {\bibinfo {volume}
  {100}},\ \bibinfo {pages} {027404} (\bibinfo {year} {2008})}\BibitemShut
  {NoStop}%
\bibitem [{\citenamefont {Shin}\ \emph {et~al.}(2015)\citenamefont {Shin},
  \citenamefont {Wolfson}, \citenamefont {Teitelbaum}, \citenamefont
  {Kandyla},\ and\ \citenamefont {Nelson}}]{Shin2015}%
  \BibitemOpen
  \bibfield  {author} {\bibinfo {author} {\bibfnamefont {T.}~\bibnamefont
  {Shin}}, \bibinfo {author} {\bibfnamefont {J.~W.}\ \bibnamefont {Wolfson}},
  \bibinfo {author} {\bibfnamefont {S.~W.}\ \bibnamefont {Teitelbaum}},
  \bibinfo {author} {\bibfnamefont {M.}~\bibnamefont {Kandyla}},\ and\ \bibinfo
  {author} {\bibfnamefont {K.~A.}\ \bibnamefont {Nelson}},\ }\href
  {https://doi.org/10.1103/physrevb.92.184302} {\bibfield  {journal} {\bibinfo
  {journal} {Physical Review B}\ }\textbf {\bibinfo {volume} {92}},\ \bibinfo
  {pages} {184302} (\bibinfo {year} {2015})}\BibitemShut {NoStop}%
\bibitem [{\citenamefont {Teitelbaum}\ \emph {et~al.}(2018)\citenamefont
  {Teitelbaum}, \citenamefont {Shin}, \citenamefont {Wolfson}, \citenamefont
  {Cheng}, \citenamefont {Porter}, \citenamefont {Kandyla},\ and\ \citenamefont
  {Nelson}}]{Teitelbaum2018}%
  \BibitemOpen
  \bibfield  {author} {\bibinfo {author} {\bibfnamefont {S.~W.}\ \bibnamefont
  {Teitelbaum}}, \bibinfo {author} {\bibfnamefont {T.}~\bibnamefont {Shin}},
  \bibinfo {author} {\bibfnamefont {J.~W.}\ \bibnamefont {Wolfson}}, \bibinfo
  {author} {\bibfnamefont {Y.-H.}\ \bibnamefont {Cheng}}, \bibinfo {author}
  {\bibfnamefont {I.~J.}\ \bibnamefont {Porter}}, \bibinfo {author}
  {\bibfnamefont {M.}~\bibnamefont {Kandyla}},\ and\ \bibinfo {author}
  {\bibfnamefont {K.~A.}\ \bibnamefont {Nelson}},\ }\href
  {https://doi.org/10.1103/physrevx.8.031081} {\bibfield  {journal} {\bibinfo
  {journal} {Physical Review X}\ }\textbf {\bibinfo {volume} {8}},\ \bibinfo
  {pages} {031081} (\bibinfo {year} {2018})}\BibitemShut {NoStop}%
\bibitem [{\citenamefont {Peierls}(1991)}]{Peierls1991}%
  \BibitemOpen
  \bibfield  {author} {\bibinfo {author} {\bibfnamefont {R.}~\bibnamefont
  {Peierls}},\ }\href
  {https://www.ebook.de/de/product/3636668/rudolf_peierls_more_surprises_in_theoretical_physics.html}
  {\emph {\bibinfo {title} {More Surprises in Theoretical Physics}}}\ (\bibinfo
   {publisher} {Princeton University Press},\ \bibinfo {year}
  {1991})\BibitemShut {NoStop}%
\bibitem [{\citenamefont {Liu}\ and\ \citenamefont {Allen}(1995)}]{Liu1995}%
  \BibitemOpen
  \bibfield  {author} {\bibinfo {author} {\bibfnamefont {Y.}~\bibnamefont
  {Liu}}\ and\ \bibinfo {author} {\bibfnamefont {R.~E.}\ \bibnamefont
  {Allen}},\ }\href {https://doi.org/10.1103/physrevb.52.1566} {\bibfield
  {journal} {\bibinfo  {journal} {Physical Review B}\ }\textbf {\bibinfo
  {volume} {52}},\ \bibinfo {pages} {1566} (\bibinfo {year}
  {1995})}\BibitemShut {NoStop}%
\bibitem [{\citenamefont {Hofmann}(2006)}]{Hofmann2006}%
  \BibitemOpen
  \bibfield  {author} {\bibinfo {author} {\bibfnamefont {P.}~\bibnamefont
  {Hofmann}},\ }\href {https://doi.org/10.1016/j.progsurf.2006.03.001}
  {\bibfield  {journal} {\bibinfo  {journal} {Progress in Surface Science}\
  }\textbf {\bibinfo {volume} {81}},\ \bibinfo {pages} {191} (\bibinfo {year}
  {2006})}\BibitemShut {NoStop}%
\bibitem [{\citenamefont {Sokolowski-Tinten}\ \emph {et~al.}(2003)\citenamefont
  {Sokolowski-Tinten}, \citenamefont {Blome}, \citenamefont {Blums},
  \citenamefont {Cavalleri}, \citenamefont {Dietrich}, \citenamefont
  {Tarasevitch}, \citenamefont {Uschmann}, \citenamefont {Förster},
  \citenamefont {Kammler}, \citenamefont {{Horn-von Hoegen}},\ and\
  \citenamefont {von~der Linde}}]{SokolowskiTinten2003}%
  \BibitemOpen
  \bibfield  {author} {\bibinfo {author} {\bibfnamefont {K.}~\bibnamefont
  {Sokolowski-Tinten}}, \bibinfo {author} {\bibfnamefont {C.}~\bibnamefont
  {Blome}}, \bibinfo {author} {\bibfnamefont {J.}~\bibnamefont {Blums}},
  \bibinfo {author} {\bibfnamefont {A.}~\bibnamefont {Cavalleri}}, \bibinfo
  {author} {\bibfnamefont {C.}~\bibnamefont {Dietrich}}, \bibinfo {author}
  {\bibfnamefont {A.}~\bibnamefont {Tarasevitch}}, \bibinfo {author}
  {\bibfnamefont {I.}~\bibnamefont {Uschmann}}, \bibinfo {author}
  {\bibfnamefont {E.}~\bibnamefont {Förster}}, \bibinfo {author}
  {\bibfnamefont {M.}~\bibnamefont {Kammler}}, \bibinfo {author} {\bibfnamefont
  {M.}~\bibnamefont {{Horn-von Hoegen}}},\ and\ \bibinfo {author}
  {\bibfnamefont {D.}~\bibnamefont {von~der Linde}},\ }\href
  {https://doi.org/10.1038/nature01490} {\bibfield  {journal} {\bibinfo
  {journal} {Nature}\ }\textbf {\bibinfo {volume} {422}},\ \bibinfo {pages}
  {287} (\bibinfo {year} {2003})}\BibitemShut {NoStop}%
\bibitem [{\citenamefont {Fritz}\ \emph {et~al.}(2007)\citenamefont {Fritz},
  \citenamefont {Reis}, \citenamefont {Adams}, \citenamefont {Akre},
  \citenamefont {Arthur}, \citenamefont {Blome}, \citenamefont {Bucksbaum},
  \citenamefont {Cavalieri}, \citenamefont {Engemann}, \citenamefont {Fahy},
  \citenamefont {Falcone}, \citenamefont {Fuoss}, \citenamefont {Gaffney},
  \citenamefont {George}, \citenamefont {Hajdu}, \citenamefont {Hertlein},
  \citenamefont {Hillyard}, \citenamefont {{Horn-von Hoegen}}, \citenamefont
  {Kammler}, \citenamefont {Kaspar}, \citenamefont {Kienberger}, \citenamefont
  {Krejcik}, \citenamefont {Lee}, \citenamefont {Lindenberg}, \citenamefont
  {McFarland}, \citenamefont {Meyer}, \citenamefont {Montagne}, \citenamefont
  {Murray}, \citenamefont {Nelson}, \citenamefont {Nicoul}, \citenamefont
  {Pahl}, \citenamefont {Rudati}, \citenamefont {Schlarb}, \citenamefont
  {Siddons}, \citenamefont {Sokolowski-Tinten}, \citenamefont {Tschentscher},
  \citenamefont {{von der Linde}},\ and\ \citenamefont {Hastings}}]{Fritz2007}%
  \BibitemOpen
  \bibfield  {author} {\bibinfo {author} {\bibfnamefont {D.~M.}\ \bibnamefont
  {Fritz}}, \bibinfo {author} {\bibfnamefont {D.~A.}\ \bibnamefont {Reis}},
  \bibinfo {author} {\bibfnamefont {B.}~\bibnamefont {Adams}}, \bibinfo
  {author} {\bibfnamefont {R.~A.}\ \bibnamefont {Akre}}, \bibinfo {author}
  {\bibfnamefont {J.}~\bibnamefont {Arthur}}, \bibinfo {author} {\bibfnamefont
  {C.}~\bibnamefont {Blome}}, \bibinfo {author} {\bibfnamefont {P.~H.}\
  \bibnamefont {Bucksbaum}}, \bibinfo {author} {\bibfnamefont {A.~L.}\
  \bibnamefont {Cavalieri}}, \bibinfo {author} {\bibfnamefont {S.}~\bibnamefont
  {Engemann}}, \bibinfo {author} {\bibfnamefont {S.}~\bibnamefont {Fahy}},
  \bibinfo {author} {\bibfnamefont {R.~W.}\ \bibnamefont {Falcone}}, \bibinfo
  {author} {\bibfnamefont {P.~H.}\ \bibnamefont {Fuoss}}, \bibinfo {author}
  {\bibfnamefont {K.~J.}\ \bibnamefont {Gaffney}}, \bibinfo {author}
  {\bibfnamefont {M.~J.}\ \bibnamefont {George}}, \bibinfo {author}
  {\bibfnamefont {J.}~\bibnamefont {Hajdu}}, \bibinfo {author} {\bibfnamefont
  {M.~P.}\ \bibnamefont {Hertlein}}, \bibinfo {author} {\bibfnamefont {P.~B.}\
  \bibnamefont {Hillyard}}, \bibinfo {author} {\bibfnamefont {M.}~\bibnamefont
  {{Horn-von Hoegen}}}, \bibinfo {author} {\bibfnamefont {M.}~\bibnamefont
  {Kammler}}, \bibinfo {author} {\bibfnamefont {J.}~\bibnamefont {Kaspar}},
  \bibinfo {author} {\bibfnamefont {R.}~\bibnamefont {Kienberger}}, \bibinfo
  {author} {\bibfnamefont {P.}~\bibnamefont {Krejcik}}, \bibinfo {author}
  {\bibfnamefont {S.~H.}\ \bibnamefont {Lee}}, \bibinfo {author} {\bibfnamefont
  {A.~M.}\ \bibnamefont {Lindenberg}}, \bibinfo {author} {\bibfnamefont
  {B.}~\bibnamefont {McFarland}}, \bibinfo {author} {\bibfnamefont
  {D.}~\bibnamefont {Meyer}}, \bibinfo {author} {\bibfnamefont
  {T.}~\bibnamefont {Montagne}}, \bibinfo {author} {\bibfnamefont {E.~D.}\
  \bibnamefont {Murray}}, \bibinfo {author} {\bibfnamefont {A.~J.}\
  \bibnamefont {Nelson}}, \bibinfo {author} {\bibfnamefont {M.}~\bibnamefont
  {Nicoul}}, \bibinfo {author} {\bibfnamefont {R.}~\bibnamefont {Pahl}},
  \bibinfo {author} {\bibfnamefont {J.}~\bibnamefont {Rudati}}, \bibinfo
  {author} {\bibfnamefont {H.}~\bibnamefont {Schlarb}}, \bibinfo {author}
  {\bibfnamefont {D.~P.}\ \bibnamefont {Siddons}}, \bibinfo {author}
  {\bibfnamefont {K.}~\bibnamefont {Sokolowski-Tinten}}, \bibinfo {author}
  {\bibfnamefont {T.}~\bibnamefont {Tschentscher}}, \bibinfo {author}
  {\bibfnamefont {D.}~\bibnamefont {{von der Linde}}},\ and\ \bibinfo {author}
  {\bibfnamefont {J.~B.}\ \bibnamefont {Hastings}},\ }\href
  {https://doi.org/10.1126/science.1135009} {\bibfield  {journal} {\bibinfo
  {journal} {Science}\ }\textbf {\bibinfo {volume} {315}},\ \bibinfo {pages}
  {633} (\bibinfo {year} {2007})}\BibitemShut {NoStop}%
\bibitem [{\citenamefont {Sciaini}\ \emph {et~al.}(2009)\citenamefont
  {Sciaini}, \citenamefont {Harb}, \citenamefont {Kruglik}, \citenamefont
  {Payer}, \citenamefont {Hebeisen}, \citenamefont {{Meyer zu Heringdorf}},
  \citenamefont {Yamaguchi}, \citenamefont {{Horn-von Hoegen}}, \citenamefont
  {Ernstorfer},\ and\ \citenamefont {Miller}}]{Sciaini2009}%
  \BibitemOpen
  \bibfield  {author} {\bibinfo {author} {\bibfnamefont {G.}~\bibnamefont
  {Sciaini}}, \bibinfo {author} {\bibfnamefont {M.}~\bibnamefont {Harb}},
  \bibinfo {author} {\bibfnamefont {S.~G.}\ \bibnamefont {Kruglik}}, \bibinfo
  {author} {\bibfnamefont {T.}~\bibnamefont {Payer}}, \bibinfo {author}
  {\bibfnamefont {C.~T.}\ \bibnamefont {Hebeisen}}, \bibinfo {author}
  {\bibfnamefont {F.-J.}\ \bibnamefont {{Meyer zu Heringdorf}}}, \bibinfo
  {author} {\bibfnamefont {M.}~\bibnamefont {Yamaguchi}}, \bibinfo {author}
  {\bibfnamefont {M.}~\bibnamefont {{Horn-von Hoegen}}}, \bibinfo {author}
  {\bibfnamefont {R.}~\bibnamefont {Ernstorfer}},\ and\ \bibinfo {author}
  {\bibfnamefont {R.~J.~D.}\ \bibnamefont {Miller}},\ }\href
  {https://doi.org/10.1038/nature07788} {\bibfield  {journal} {\bibinfo
  {journal} {Nature}\ }\textbf {\bibinfo {volume} {458}},\ \bibinfo {pages}
  {56} (\bibinfo {year} {2009})}\BibitemShut {NoStop}%
\bibitem [{\citenamefont {Timrov}\ \emph {et~al.}(2012)\citenamefont {Timrov},
  \citenamefont {Kampfrath}, \citenamefont {Faure}, \citenamefont {Vast},
  \citenamefont {Ast}, \citenamefont {Frischkorn}, \citenamefont {Wolf},
  \citenamefont {Gava},\ and\ \citenamefont {Perfetti}}]{Timrov2012}%
  \BibitemOpen
  \bibfield  {author} {\bibinfo {author} {\bibfnamefont {I.}~\bibnamefont
  {Timrov}}, \bibinfo {author} {\bibfnamefont {T.}~\bibnamefont {Kampfrath}},
  \bibinfo {author} {\bibfnamefont {J.}~\bibnamefont {Faure}}, \bibinfo
  {author} {\bibfnamefont {N.}~\bibnamefont {Vast}}, \bibinfo {author}
  {\bibfnamefont {C.~R.}\ \bibnamefont {Ast}}, \bibinfo {author} {\bibfnamefont
  {C.}~\bibnamefont {Frischkorn}}, \bibinfo {author} {\bibfnamefont
  {M.}~\bibnamefont {Wolf}}, \bibinfo {author} {\bibfnamefont {P.}~\bibnamefont
  {Gava}},\ and\ \bibinfo {author} {\bibfnamefont {L.}~\bibnamefont
  {Perfetti}},\ }\href {https://doi.org/10.1103/physrevb.85.155139} {\bibfield
  {journal} {\bibinfo  {journal} {Physical Review B}\ }\textbf {\bibinfo
  {volume} {85}},\ \bibinfo {pages} {155139} (\bibinfo {year}
  {2012})}\BibitemShut {NoStop}%
\bibitem [{\citenamefont {Kammler}\ and\ \citenamefont {{Horn-von
  Hoegen}}(2005)}]{Kammler2005}%
  \BibitemOpen
  \bibfield  {author} {\bibinfo {author} {\bibfnamefont {M.}~\bibnamefont
  {Kammler}}\ and\ \bibinfo {author} {\bibfnamefont {M.}~\bibnamefont
  {{Horn-von Hoegen}}},\ }\href {https://doi.org/10.1016/j.susc.2004.11.033}
  {\bibfield  {journal} {\bibinfo  {journal} {Surface Science}\ }\textbf
  {\bibinfo {volume} {576}},\ \bibinfo {pages} {56} (\bibinfo {year}
  {2005})}\BibitemShut {NoStop}%
\bibitem [{\citenamefont {Payer}\ \emph {et~al.}(2012)\citenamefont {Payer},
  \citenamefont {Klein}, \citenamefont {Acet}, \citenamefont {Ney},
  \citenamefont {Kammler}, \citenamefont {{Meyer zu Heringdorf}},\ and\
  \citenamefont {{Horn-von Hoegen}}}]{Payer2012}%
  \BibitemOpen
  \bibfield  {author} {\bibinfo {author} {\bibfnamefont {T.}~\bibnamefont
  {Payer}}, \bibinfo {author} {\bibfnamefont {C.}~\bibnamefont {Klein}},
  \bibinfo {author} {\bibfnamefont {M.}~\bibnamefont {Acet}}, \bibinfo {author}
  {\bibfnamefont {V.}~\bibnamefont {Ney}}, \bibinfo {author} {\bibfnamefont
  {M.}~\bibnamefont {Kammler}}, \bibinfo {author} {\bibfnamefont {F.-J.}\
  \bibnamefont {{Meyer zu Heringdorf}}},\ and\ \bibinfo {author} {\bibfnamefont
  {M.}~\bibnamefont {{Horn-von Hoegen}}},\ }\href
  {https://doi.org/10.1016/j.tsf.2012.06.004} {\bibfield  {journal} {\bibinfo
  {journal} {Thin Solid Films}\ }\textbf {\bibinfo {volume} {520}},\ \bibinfo
  {pages} {6905} (\bibinfo {year} {2012})}\BibitemShut {NoStop}%
\bibitem [{\citenamefont {Toudert}\ \emph {et~al.}(2017)\citenamefont
  {Toudert}, \citenamefont {Serna}, \citenamefont {Camps}, \citenamefont
  {Wojcik}, \citenamefont {Mascher}, \citenamefont {Rebollar},\ and\
  \citenamefont {Ezquerra}}]{Toudert2017}%
  \BibitemOpen
  \bibfield  {author} {\bibinfo {author} {\bibfnamefont {J.}~\bibnamefont
  {Toudert}}, \bibinfo {author} {\bibfnamefont {R.}~\bibnamefont {Serna}},
  \bibinfo {author} {\bibfnamefont {I.}~\bibnamefont {Camps}}, \bibinfo
  {author} {\bibfnamefont {J.}~\bibnamefont {Wojcik}}, \bibinfo {author}
  {\bibfnamefont {P.}~\bibnamefont {Mascher}}, \bibinfo {author} {\bibfnamefont
  {E.}~\bibnamefont {Rebollar}},\ and\ \bibinfo {author} {\bibfnamefont
  {T.~A.}\ \bibnamefont {Ezquerra}},\ }\href
  {https://doi.org/10.1021/acs.jpcc.6b10331} {\bibfield  {journal} {\bibinfo
  {journal} {The Journal of Physical Chemistry C}\ }\textbf {\bibinfo {volume}
  {121}},\ \bibinfo {pages} {3511} (\bibinfo {year} {2017})}\BibitemShut
  {NoStop}%
\bibitem [{\citenamefont {Zeiger}\ \emph {et~al.}(1992)\citenamefont {Zeiger},
  \citenamefont {Vidal}, \citenamefont {Cheng}, \citenamefont {Ippen},
  \citenamefont {Dresselhaus},\ and\ \citenamefont {Dresselhaus}}]{Zeiger1992}%
  \BibitemOpen
  \bibfield  {author} {\bibinfo {author} {\bibfnamefont {H.~J.}\ \bibnamefont
  {Zeiger}}, \bibinfo {author} {\bibfnamefont {J.}~\bibnamefont {Vidal}},
  \bibinfo {author} {\bibfnamefont {T.~K.}\ \bibnamefont {Cheng}}, \bibinfo
  {author} {\bibfnamefont {E.~P.}\ \bibnamefont {Ippen}}, \bibinfo {author}
  {\bibfnamefont {G.}~\bibnamefont {Dresselhaus}},\ and\ \bibinfo {author}
  {\bibfnamefont {M.~S.}\ \bibnamefont {Dresselhaus}},\ }\href
  {https://doi.org/10.1103/physrevb.45.768} {\bibfield  {journal} {\bibinfo
  {journal} {Physical Review B}\ }\textbf {\bibinfo {volume} {45}},\ \bibinfo
  {pages} {768} (\bibinfo {year} {1992})}\BibitemShut {NoStop}%
\bibitem [{\citenamefont {Misochko}\ \emph {et~al.}(2004)\citenamefont
  {Misochko}, \citenamefont {Hase}, \citenamefont {Ishioka},\ and\
  \citenamefont {Kitajima}}]{Misochko2004}%
  \BibitemOpen
  \bibfield  {author} {\bibinfo {author} {\bibfnamefont {O.~V.}\ \bibnamefont
  {Misochko}}, \bibinfo {author} {\bibfnamefont {M.}~\bibnamefont {Hase}},
  \bibinfo {author} {\bibfnamefont {K.}~\bibnamefont {Ishioka}},\ and\ \bibinfo
  {author} {\bibfnamefont {M.}~\bibnamefont {Kitajima}},\ }\href
  {https://doi.org/10.1103/physrevlett.92.197401} {\bibfield  {journal}
  {\bibinfo  {journal} {Physical Review Letters}\ }\textbf {\bibinfo {volume}
  {92}},\ \bibinfo {pages} {197401} (\bibinfo {year} {2004})}\BibitemShut
  {NoStop}%
\bibitem [{\citenamefont {Ishioka}\ \emph {et~al.}(2006)\citenamefont
  {Ishioka}, \citenamefont {Kitajima},\ and\ \citenamefont
  {Misochko}}]{Ishioka2006}%
  \BibitemOpen
  \bibfield  {author} {\bibinfo {author} {\bibfnamefont {K.}~\bibnamefont
  {Ishioka}}, \bibinfo {author} {\bibfnamefont {M.}~\bibnamefont {Kitajima}},\
  and\ \bibinfo {author} {\bibfnamefont {O.~V.}\ \bibnamefont {Misochko}},\
  }\href {https://doi.org/10.1063/1.2363746} {\bibfield  {journal} {\bibinfo
  {journal} {Journal of Applied Physics}\ }\textbf {\bibinfo {volume} {100}},\
  \bibinfo {pages} {093501} (\bibinfo {year} {2006})}\BibitemShut {NoStop}%
\bibitem [{\citenamefont {Shin}(2018)}]{Shin2018}%
  \BibitemOpen
  \bibfield  {author} {\bibinfo {author} {\bibfnamefont {T.}~\bibnamefont
  {Shin}},\ }\href {https://doi.org/10.1016/j.tsf.2018.09.037} {\bibfield
  {journal} {\bibinfo  {journal} {Thin Solid Films}\ }\textbf {\bibinfo
  {volume} {666}},\ \bibinfo {pages} {108} (\bibinfo {year}
  {2018})}\BibitemShut {NoStop}%
\bibitem [{\citenamefont {Thiemann}\ \emph {et~al.}(2022)\citenamefont
  {Thiemann}, \citenamefont {Sciaini}, \citenamefont {Kassen}, \citenamefont
  {Hagemann}, \citenamefont {{Meyer zu Heringdorf}},\ and\ \citenamefont
  {{Horn-von Hoegen}}}]{Thiemann2022}%
  \BibitemOpen
  \bibfield  {author} {\bibinfo {author} {\bibfnamefont {F.}~\bibnamefont
  {Thiemann}}, \bibinfo {author} {\bibfnamefont {G.}~\bibnamefont {Sciaini}},
  \bibinfo {author} {\bibfnamefont {A.}~\bibnamefont {Kassen}}, \bibinfo
  {author} {\bibfnamefont {U.}~\bibnamefont {Hagemann}}, \bibinfo {author}
  {\bibfnamefont {F.}~\bibnamefont {{Meyer zu Heringdorf}}},\ and\ \bibinfo
  {author} {\bibfnamefont {M.}~\bibnamefont {{Horn-von Hoegen}}},\ }\href
  {https://doi.org/10.1103/physrevb.106.014315} {\bibfield  {journal} {\bibinfo
   {journal} {Physical Review B}\ }\textbf {\bibinfo {volume} {106}},\ \bibinfo
  {pages} {014315} (\bibinfo {year} {2022})}\BibitemShut {NoStop}%
\bibitem [{\citenamefont {Jnawali}\ \emph {et~al.}(2021)\citenamefont
  {Jnawali}, \citenamefont {Boschetto}, \citenamefont {Malard}, \citenamefont
  {Heinz}, \citenamefont {Sciaini}, \citenamefont {Thiemann}, \citenamefont
  {Payer}, \citenamefont {Kremeyer}, \citenamefont {{Meyer zu Heringdorf}},\
  and\ \citenamefont {{Horn-von Hoegen}}}]{Jnawali2021}%
  \BibitemOpen
  \bibfield  {author} {\bibinfo {author} {\bibfnamefont {G.}~\bibnamefont
  {Jnawali}}, \bibinfo {author} {\bibfnamefont {D.}~\bibnamefont {Boschetto}},
  \bibinfo {author} {\bibfnamefont {L.~M.}\ \bibnamefont {Malard}}, \bibinfo
  {author} {\bibfnamefont {T.~F.}\ \bibnamefont {Heinz}}, \bibinfo {author}
  {\bibfnamefont {G.}~\bibnamefont {Sciaini}}, \bibinfo {author} {\bibfnamefont
  {F.}~\bibnamefont {Thiemann}}, \bibinfo {author} {\bibfnamefont
  {T.}~\bibnamefont {Payer}}, \bibinfo {author} {\bibfnamefont
  {L.}~\bibnamefont {Kremeyer}}, \bibinfo {author} {\bibfnamefont {F.-J.}\
  \bibnamefont {{Meyer zu Heringdorf}}},\ and\ \bibinfo {author} {\bibfnamefont
  {M.}~\bibnamefont {{Horn-von Hoegen}}},\ }\href
  {https://doi.org/10.1063/5.0056813} {\bibfield  {journal} {\bibinfo
  {journal} {Applied Physics Letters}\ }\textbf {\bibinfo {volume} {119}},\
  \bibinfo {pages} {091601} (\bibinfo {year} {2021})}\BibitemShut {NoStop}%
\bibitem [{\citenamefont {Hanisch-Blicharski}\ \emph
  {et~al.}(2021)\citenamefont {Hanisch-Blicharski}, \citenamefont {Tinnemann},
  \citenamefont {Wall}, \citenamefont {Thiemann}, \citenamefont {Groven},
  \citenamefont {Fortmann}, \citenamefont {Tajik}, \citenamefont {Brand},
  \citenamefont {Frost}, \citenamefont {von Hoegen},\ and\ \citenamefont
  {{Horn-von Hoegen}}}]{HanischBlicharski2021}%
  \BibitemOpen
  \bibfield  {author} {\bibinfo {author} {\bibfnamefont {A.}~\bibnamefont
  {Hanisch-Blicharski}}, \bibinfo {author} {\bibfnamefont {V.}~\bibnamefont
  {Tinnemann}}, \bibinfo {author} {\bibfnamefont {S.}~\bibnamefont {Wall}},
  \bibinfo {author} {\bibfnamefont {F.}~\bibnamefont {Thiemann}}, \bibinfo
  {author} {\bibfnamefont {T.}~\bibnamefont {Groven}}, \bibinfo {author}
  {\bibfnamefont {J.}~\bibnamefont {Fortmann}}, \bibinfo {author}
  {\bibfnamefont {M.}~\bibnamefont {Tajik}}, \bibinfo {author} {\bibfnamefont
  {C.}~\bibnamefont {Brand}}, \bibinfo {author} {\bibfnamefont {B.-O.}\
  \bibnamefont {Frost}}, \bibinfo {author} {\bibfnamefont {A.}~\bibnamefont
  {von Hoegen}},\ and\ \bibinfo {author} {\bibfnamefont {M.}~\bibnamefont
  {{Horn-von Hoegen}}},\ }\href {https://doi.org/10.1021/acs.nanolett.1c01665}
  {\bibfield  {journal} {\bibinfo  {journal} {Nano Letters}\ }\textbf {\bibinfo
  {volume} {21}},\ \bibinfo {pages} {7145} (\bibinfo {year}
  {2021})}\BibitemShut {NoStop}%
\bibitem [{\citenamefont {He}\ \emph {et~al.}(2020)\citenamefont {He},
  \citenamefont {Walker}, \citenamefont {Zhou}, \citenamefont {Montano},
  \citenamefont {Bank},\ and\ \citenamefont {Wang}}]{He2020}%
  \BibitemOpen
  \bibfield  {author} {\bibinfo {author} {\bibfnamefont {F.}~\bibnamefont
  {He}}, \bibinfo {author} {\bibfnamefont {E.~S.}\ \bibnamefont {Walker}},
  \bibinfo {author} {\bibfnamefont {Y.}~\bibnamefont {Zhou}}, \bibinfo {author}
  {\bibfnamefont {R.~D.}\ \bibnamefont {Montano}}, \bibinfo {author}
  {\bibfnamefont {S.~R.}\ \bibnamefont {Bank}},\ and\ \bibinfo {author}
  {\bibfnamefont {Y.}~\bibnamefont {Wang}},\ }\href
  {https://doi.org/10.1063/5.0016793} {\bibfield  {journal} {\bibinfo
  {journal} {Applied Physics Letters}\ }\textbf {\bibinfo {volume} {117}},\
  \bibinfo {pages} {073103} (\bibinfo {year} {2020})}\BibitemShut {NoStop}%
\bibitem [{\citenamefont {Murray}\ \emph {et~al.}(2005)\citenamefont {Murray},
  \citenamefont {Fritz}, \citenamefont {Wahlstrand}, \citenamefont {Fahy},\
  and\ \citenamefont {Reis}}]{Murray2005}%
  \BibitemOpen
  \bibfield  {author} {\bibinfo {author} {\bibfnamefont {{\'{E}}.~D.}\
  \bibnamefont {Murray}}, \bibinfo {author} {\bibfnamefont {D.~M.}\
  \bibnamefont {Fritz}}, \bibinfo {author} {\bibfnamefont {J.~K.}\ \bibnamefont
  {Wahlstrand}}, \bibinfo {author} {\bibfnamefont {S.}~\bibnamefont {Fahy}},\
  and\ \bibinfo {author} {\bibfnamefont {D.~A.}\ \bibnamefont {Reis}},\ }\href
  {https://doi.org/10.1103/physrevb.72.060301} {\bibfield  {journal} {\bibinfo
  {journal} {Physical Review B}\ }\textbf {\bibinfo {volume} {72}},\ \bibinfo
  {pages} {060301} (\bibinfo {year} {2005})}\BibitemShut {NoStop}%
\bibitem [{\citenamefont {Boschini}\ \emph {et~al.}(2015)\citenamefont
  {Boschini}, \citenamefont {Hedayat}, \citenamefont {Piovera}, \citenamefont
  {Dallera}, \citenamefont {Gupta},\ and\ \citenamefont
  {Carpene}}]{Boschini2015}%
  \BibitemOpen
  \bibfield  {author} {\bibinfo {author} {\bibfnamefont {F.}~\bibnamefont
  {Boschini}}, \bibinfo {author} {\bibfnamefont {H.}~\bibnamefont {Hedayat}},
  \bibinfo {author} {\bibfnamefont {C.}~\bibnamefont {Piovera}}, \bibinfo
  {author} {\bibfnamefont {C.}~\bibnamefont {Dallera}}, \bibinfo {author}
  {\bibfnamefont {A.}~\bibnamefont {Gupta}},\ and\ \bibinfo {author}
  {\bibfnamefont {E.}~\bibnamefont {Carpene}},\ }\href
  {https://doi.org/10.1063/1.4906756} {\bibfield  {journal} {\bibinfo
  {journal} {Review of Scientific Instruments}\ }\textbf {\bibinfo {volume}
  {86}},\ \bibinfo {pages} {013909} (\bibinfo {year} {2015})}\BibitemShut
  {NoStop}%
\bibitem [{\citenamefont {Richter}\ \emph {et~al.}(2021)\citenamefont
  {Richter}, \citenamefont {Rebarz}, \citenamefont {Herrfurth}, \citenamefont
  {Espinoza}, \citenamefont {Schmidt-Grund},\ and\ \citenamefont
  {Andreasson}}]{Richter2021}%
  \BibitemOpen
  \bibfield  {author} {\bibinfo {author} {\bibfnamefont {S.}~\bibnamefont
  {Richter}}, \bibinfo {author} {\bibfnamefont {M.}~\bibnamefont {Rebarz}},
  \bibinfo {author} {\bibfnamefont {O.}~\bibnamefont {Herrfurth}}, \bibinfo
  {author} {\bibfnamefont {S.}~\bibnamefont {Espinoza}}, \bibinfo {author}
  {\bibfnamefont {R.}~\bibnamefont {Schmidt-Grund}},\ and\ \bibinfo {author}
  {\bibfnamefont {J.}~\bibnamefont {Andreasson}},\ }\href
  {https://doi.org/10.1063/5.0027219} {\bibfield  {journal} {\bibinfo
  {journal} {Review of Scientific Instruments}\ }\textbf {\bibinfo {volume}
  {92}},\ \bibinfo {pages} {033104} (\bibinfo {year} {2021})}\BibitemShut
  {NoStop}%
\bibitem [{\citenamefont {Roeser}\ \emph {et~al.}(2003)\citenamefont {Roeser},
  \citenamefont {Kim}, \citenamefont {Callan}, \citenamefont {Huang},
  \citenamefont {Glezer}, \citenamefont {Siegal},\ and\ \citenamefont
  {Mazur}}]{Roeser2003}%
  \BibitemOpen
  \bibfield  {author} {\bibinfo {author} {\bibfnamefont {C.~A.~D.}\
  \bibnamefont {Roeser}}, \bibinfo {author} {\bibfnamefont {A.~M.-T.}\
  \bibnamefont {Kim}}, \bibinfo {author} {\bibfnamefont {J.~P.}\ \bibnamefont
  {Callan}}, \bibinfo {author} {\bibfnamefont {L.}~\bibnamefont {Huang}},
  \bibinfo {author} {\bibfnamefont {E.~N.}\ \bibnamefont {Glezer}}, \bibinfo
  {author} {\bibfnamefont {Y.}~\bibnamefont {Siegal}},\ and\ \bibinfo {author}
  {\bibfnamefont {E.}~\bibnamefont {Mazur}},\ }\href
  {https://doi.org/10.1063/1.1582383} {\bibfield  {journal} {\bibinfo
  {journal} {Review of Scientific Instruments}\ }\textbf {\bibinfo {volume}
  {74}},\ \bibinfo {pages} {3413} (\bibinfo {year} {2003})}\BibitemShut
  {NoStop}%
\bibitem [{\citenamefont {Hricovini}\ \emph {et~al.}(1992)\citenamefont
  {Hricovini}, \citenamefont {Lay}, \citenamefont {Kahn}, \citenamefont
  {Taleb-Ibrahimi},\ and\ \citenamefont {Bonnet}}]{Hricovini1992}%
  \BibitemOpen
  \bibfield  {author} {\bibinfo {author} {\bibfnamefont {K.}~\bibnamefont
  {Hricovini}}, \bibinfo {author} {\bibfnamefont {G.~L.}\ \bibnamefont {Lay}},
  \bibinfo {author} {\bibfnamefont {A.}~\bibnamefont {Kahn}}, \bibinfo {author}
  {\bibfnamefont {A.}~\bibnamefont {Taleb-Ibrahimi}},\ and\ \bibinfo {author}
  {\bibfnamefont {J.}~\bibnamefont {Bonnet}},\ }\href
  {https://doi.org/10.1016/0169-4332(92)90244-r} {\bibfield  {journal}
  {\bibinfo  {journal} {Applied Surface Science}\ }\textbf {\bibinfo {volume}
  {56-58}},\ \bibinfo {pages} {259} (\bibinfo {year} {1992})}\BibitemShut
  {NoStop}%
\bibitem [{\citenamefont {{De Renzi}}\ \emph {et~al.}(1993)\citenamefont {{De
  Renzi}}, \citenamefont {Betti},\ and\ \citenamefont {Mariani}}]{Renzi1993}%
  \BibitemOpen
  \bibfield  {author} {\bibinfo {author} {\bibfnamefont {V.}~\bibnamefont {{De
  Renzi}}}, \bibinfo {author} {\bibfnamefont {M.~G.}\ \bibnamefont {Betti}},\
  and\ \bibinfo {author} {\bibfnamefont {C.}~\bibnamefont {Mariani}},\ }\href
  {https://doi.org/10.1103/physrevb.48.4767} {\bibfield  {journal} {\bibinfo
  {journal} {Physical Review B}\ }\textbf {\bibinfo {volume} {48}},\ \bibinfo
  {pages} {4767} (\bibinfo {year} {1993})}\BibitemShut {NoStop}%
\bibitem [{\citenamefont {Katsidis}\ and\ \citenamefont
  {Siapkas}(2002)}]{Katsidis2002}%
  \BibitemOpen
  \bibfield  {author} {\bibinfo {author} {\bibfnamefont {C.~C.}\ \bibnamefont
  {Katsidis}}\ and\ \bibinfo {author} {\bibfnamefont {D.~I.}\ \bibnamefont
  {Siapkas}},\ }\href {https://doi.org/10.1364/ao.41.003978} {\bibfield
  {journal} {\bibinfo  {journal} {Applied Optics}\ }\textbf {\bibinfo {volume}
  {41}},\ \bibinfo {pages} {3978} (\bibinfo {year} {2002})}\BibitemShut
  {NoStop}%
\bibitem [{\citenamefont {Aguilera}\ \emph {et~al.}(2015)\citenamefont
  {Aguilera}, \citenamefont {Friedrich},\ and\ \citenamefont
  {Blügel}}]{Aguilera2015}%
  \BibitemOpen
  \bibfield  {author} {\bibinfo {author} {\bibfnamefont {I.}~\bibnamefont
  {Aguilera}}, \bibinfo {author} {\bibfnamefont {C.}~\bibnamefont
  {Friedrich}},\ and\ \bibinfo {author} {\bibfnamefont {S.}~\bibnamefont
  {Blügel}},\ }\href {https://doi.org/10.1103/physrevb.91.125129} {\bibfield
  {journal} {\bibinfo  {journal} {Physical Review B}\ }\textbf {\bibinfo
  {volume} {91}},\ \bibinfo {pages} {125129} (\bibinfo {year}
  {2015})}\BibitemShut {NoStop}%
\bibitem [{\citenamefont {Cardona}\ and\ \citenamefont
  {Greenaway}(1964)}]{Cardona1964}%
  \BibitemOpen
  \bibfield  {author} {\bibinfo {author} {\bibfnamefont {M.}~\bibnamefont
  {Cardona}}\ and\ \bibinfo {author} {\bibfnamefont {D.~L.}\ \bibnamefont
  {Greenaway}},\ }\href {https://doi.org/10.1103/physrev.133.a1685} {\bibfield
  {journal} {\bibinfo  {journal} {Physical Review}\ }\textbf {\bibinfo {volume}
  {133}},\ \bibinfo {pages} {A1685} (\bibinfo {year} {1964})}\BibitemShut
  {NoStop}%
\bibitem [{\citenamefont {Hunderi}(1975)}]{Hunderi1975}%
  \BibitemOpen
  \bibfield  {author} {\bibinfo {author} {\bibfnamefont {O.}~\bibnamefont
  {Hunderi}},\ }\href {https://doi.org/10.1088/0305-4608/5/11/034} {\bibfield
  {journal} {\bibinfo  {journal} {Journal of Physics F: Metal Physics}\
  }\textbf {\bibinfo {volume} {5}},\ \bibinfo {pages} {2214} (\bibinfo {year}
  {1975})}\BibitemShut {NoStop}%
\bibitem [{\citenamefont {Liu}\ \emph {et~al.}(2020)\citenamefont {Liu},
  \citenamefont {Yang}, \citenamefont {Chen}, \citenamefont {Chen},
  \citenamefont {Guo}, \citenamefont {Saito}, \citenamefont {Li},\ and\
  \citenamefont {Li}}]{Liu2020}%
  \BibitemOpen
  \bibfield  {author} {\bibinfo {author} {\bibfnamefont {H.-L.}\ \bibnamefont
  {Liu}}, \bibinfo {author} {\bibfnamefont {T.}~\bibnamefont {Yang}}, \bibinfo
  {author} {\bibfnamefont {J.-H.}\ \bibnamefont {Chen}}, \bibinfo {author}
  {\bibfnamefont {H.-W.}\ \bibnamefont {Chen}}, \bibinfo {author}
  {\bibfnamefont {H.}~\bibnamefont {Guo}}, \bibinfo {author} {\bibfnamefont
  {R.}~\bibnamefont {Saito}}, \bibinfo {author} {\bibfnamefont {M.-Y.}\
  \bibnamefont {Li}},\ and\ \bibinfo {author} {\bibfnamefont {L.-J.}\
  \bibnamefont {Li}},\ }\bibfield  {journal} {\bibinfo  {journal} {Scientific
  Reports}\ }\textbf {\bibinfo {volume} {10}},\ \href
  {https://doi.org/10.1038/s41598-020-71808-y} {10.1038/s41598-020-71808-y}
  (\bibinfo {year} {2020})\BibitemShut {NoStop}%
\bibitem [{\citenamefont {Han}\ \emph {et~al.}(2023)\citenamefont {Han},
  \citenamefont {Lee}, \citenamefont {Song}, \citenamefont {Wang},
  \citenamefont {Bermel},\ and\ \citenamefont {Ruan}}]{Han2023}%
  \BibitemOpen
  \bibfield  {author} {\bibinfo {author} {\bibfnamefont {Z.}~\bibnamefont
  {Han}}, \bibinfo {author} {\bibfnamefont {C.}~\bibnamefont {Lee}}, \bibinfo
  {author} {\bibfnamefont {J.}~\bibnamefont {Song}}, \bibinfo {author}
  {\bibfnamefont {H.}~\bibnamefont {Wang}}, \bibinfo {author} {\bibfnamefont
  {P.}~\bibnamefont {Bermel}},\ and\ \bibinfo {author} {\bibfnamefont
  {X.}~\bibnamefont {Ruan}},\ }\href
  {https://doi.org/10.1103/physrevb.107.l201202} {\bibfield  {journal}
  {\bibinfo  {journal} {Physical Review B}\ }\textbf {\bibinfo {volume}
  {107}},\ \bibinfo {pages} {l201202} (\bibinfo {year} {2023})}\BibitemShut
  {NoStop}%
\bibitem [{\citenamefont {Smejkal}\ \emph {et~al.}(2022)\citenamefont
  {Smejkal}, \citenamefont {Trovatello}, \citenamefont {Li}, \citenamefont
  {Conte}, \citenamefont {Marini}, \citenamefont {Zhu}, \citenamefont
  {Cerullo},\ and\ \citenamefont {Libisch}}]{Smejkal2022}%
  \BibitemOpen
  \bibfield  {author} {\bibinfo {author} {\bibfnamefont {V.}~\bibnamefont
  {Smejkal}}, \bibinfo {author} {\bibfnamefont {C.}~\bibnamefont {Trovatello}},
  \bibinfo {author} {\bibfnamefont {Q.}~\bibnamefont {Li}}, \bibinfo {author}
  {\bibfnamefont {S.~D.}\ \bibnamefont {Conte}}, \bibinfo {author}
  {\bibfnamefont {A.}~\bibnamefont {Marini}}, \bibinfo {author} {\bibfnamefont
  {X.}~\bibnamefont {Zhu}}, \bibinfo {author} {\bibfnamefont {G.}~\bibnamefont
  {Cerullo}},\ and\ \bibinfo {author} {\bibfnamefont {F.}~\bibnamefont
  {Libisch}},\ }\href {https://doi.org/10.1364/oe.479518} {\bibfield  {journal}
  {\bibinfo  {journal} {Optics Express}\ }\textbf {\bibinfo {volume} {31}},\
  \bibinfo {pages} {107} (\bibinfo {year} {2022})}\BibitemShut {NoStop}%
\bibitem [{\citenamefont {Trovatello}\ \emph {et~al.}(2022)\citenamefont
  {Trovatello}, \citenamefont {Katsch}, \citenamefont {Li}, \citenamefont
  {Zhu}, \citenamefont {Knorr}, \citenamefont {Cerullo},\ and\ \citenamefont
  {Conte}}]{Trovatello2022}%
  \BibitemOpen
  \bibfield  {author} {\bibinfo {author} {\bibfnamefont {C.}~\bibnamefont
  {Trovatello}}, \bibinfo {author} {\bibfnamefont {F.}~\bibnamefont {Katsch}},
  \bibinfo {author} {\bibfnamefont {Q.}~\bibnamefont {Li}}, \bibinfo {author}
  {\bibfnamefont {X.}~\bibnamefont {Zhu}}, \bibinfo {author} {\bibfnamefont
  {A.}~\bibnamefont {Knorr}}, \bibinfo {author} {\bibfnamefont
  {G.}~\bibnamefont {Cerullo}},\ and\ \bibinfo {author} {\bibfnamefont {S.~D.}\
  \bibnamefont {Conte}},\ }\href {https://doi.org/10.1021/acs.nanolett.2c01309}
  {\bibfield  {journal} {\bibinfo  {journal} {Nano Letters}\ }\textbf {\bibinfo
  {volume} {22}},\ \bibinfo {pages} {5322} (\bibinfo {year}
  {2022})}\BibitemShut {NoStop}%
\bibitem [{\citenamefont {Vi{\~{n}}a}\ \emph {et~al.}(1984)\citenamefont
  {Vi{\~{n}}a}, \citenamefont {Logothetidis},\ and\ \citenamefont
  {Cardona}}]{Vina1984}%
  \BibitemOpen
  \bibfield  {author} {\bibinfo {author} {\bibfnamefont {L.}~\bibnamefont
  {Vi{\~{n}}a}}, \bibinfo {author} {\bibfnamefont {S.}~\bibnamefont
  {Logothetidis}},\ and\ \bibinfo {author} {\bibfnamefont {M.}~\bibnamefont
  {Cardona}},\ }\href {https://doi.org/10.1103/physrevb.30.1979} {\bibfield
  {journal} {\bibinfo  {journal} {Physical Review B}\ }\textbf {\bibinfo
  {volume} {30}},\ \bibinfo {pages} {1979} (\bibinfo {year}
  {1984})}\BibitemShut {NoStop}%
\bibitem [{\citenamefont {Lautenschlager}\ \emph
  {et~al.}(1987{\natexlab{a}})\citenamefont {Lautenschlager}, \citenamefont
  {Garriga}, \citenamefont {Vina},\ and\ \citenamefont
  {Cardona}}]{Lautenschlager1987}%
  \BibitemOpen
  \bibfield  {author} {\bibinfo {author} {\bibfnamefont {P.}~\bibnamefont
  {Lautenschlager}}, \bibinfo {author} {\bibfnamefont {M.}~\bibnamefont
  {Garriga}}, \bibinfo {author} {\bibfnamefont {L.}~\bibnamefont {Vina}},\ and\
  \bibinfo {author} {\bibfnamefont {M.}~\bibnamefont {Cardona}},\ }\href
  {https://doi.org/10.1103/physrevb.36.4821} {\bibfield  {journal} {\bibinfo
  {journal} {Physical Review B}\ }\textbf {\bibinfo {volume} {36}},\ \bibinfo
  {pages} {4821} (\bibinfo {year} {1987}{\natexlab{a}})}\BibitemShut {NoStop}%
\bibitem [{\citenamefont {Lautenschlager}\ \emph
  {et~al.}(1987{\natexlab{b}})\citenamefont {Lautenschlager}, \citenamefont
  {Garriga}, \citenamefont {Logothetidis},\ and\ \citenamefont
  {Cardona}}]{Lautenschlager1987a}%
  \BibitemOpen
  \bibfield  {author} {\bibinfo {author} {\bibfnamefont {P.}~\bibnamefont
  {Lautenschlager}}, \bibinfo {author} {\bibfnamefont {M.}~\bibnamefont
  {Garriga}}, \bibinfo {author} {\bibfnamefont {S.}~\bibnamefont
  {Logothetidis}},\ and\ \bibinfo {author} {\bibfnamefont {M.}~\bibnamefont
  {Cardona}},\ }\href {https://doi.org/10.1103/physrevb.35.9174} {\bibfield
  {journal} {\bibinfo  {journal} {Physical Review B}\ }\textbf {\bibinfo
  {volume} {35}},\ \bibinfo {pages} {9174} (\bibinfo {year}
  {1987}{\natexlab{b}})}\BibitemShut {NoStop}%
\bibitem [{\citenamefont {Shkrebtii}\ \emph {et~al.}(2010)\citenamefont
  {Shkrebtii}, \citenamefont {Ibrahim}, \citenamefont {Teatro}, \citenamefont
  {Richter}, \citenamefont {Lee},\ and\ \citenamefont
  {Henderson}}]{Shkrebtii2010}%
  \BibitemOpen
  \bibfield  {author} {\bibinfo {author} {\bibfnamefont {A.~I.}\ \bibnamefont
  {Shkrebtii}}, \bibinfo {author} {\bibfnamefont {Z.~A.}\ \bibnamefont
  {Ibrahim}}, \bibinfo {author} {\bibfnamefont {T.}~\bibnamefont {Teatro}},
  \bibinfo {author} {\bibfnamefont {W.}~\bibnamefont {Richter}}, \bibinfo
  {author} {\bibfnamefont {M.~J.}\ \bibnamefont {Lee}},\ and\ \bibinfo {author}
  {\bibfnamefont {L.}~\bibnamefont {Henderson}},\ }\href
  {https://doi.org/10.1002/pssb.200983942} {\bibfield  {journal} {\bibinfo
  {journal} {physica status solidi (b)}\ }\textbf {\bibinfo {volume} {247}},\
  \bibinfo {pages} {1881} (\bibinfo {year} {2010})}\BibitemShut {NoStop}%
\bibitem [{\citenamefont {Faure}\ \emph {et~al.}(2013)\citenamefont {Faure},
  \citenamefont {Mauchain}, \citenamefont {Papalazarou}, \citenamefont {Marsi},
  \citenamefont {Boschetto}, \citenamefont {Timrov}, \citenamefont {Vast},
  \citenamefont {Ohtsubo}, \citenamefont {Arnaud},\ and\ \citenamefont
  {Perfetti}}]{Faure2013}%
  \BibitemOpen
  \bibfield  {author} {\bibinfo {author} {\bibfnamefont {J.}~\bibnamefont
  {Faure}}, \bibinfo {author} {\bibfnamefont {J.}~\bibnamefont {Mauchain}},
  \bibinfo {author} {\bibfnamefont {E.}~\bibnamefont {Papalazarou}}, \bibinfo
  {author} {\bibfnamefont {M.}~\bibnamefont {Marsi}}, \bibinfo {author}
  {\bibfnamefont {D.}~\bibnamefont {Boschetto}}, \bibinfo {author}
  {\bibfnamefont {I.}~\bibnamefont {Timrov}}, \bibinfo {author} {\bibfnamefont
  {N.}~\bibnamefont {Vast}}, \bibinfo {author} {\bibfnamefont {Y.}~\bibnamefont
  {Ohtsubo}}, \bibinfo {author} {\bibfnamefont {B.}~\bibnamefont {Arnaud}},\
  and\ \bibinfo {author} {\bibfnamefont {L.}~\bibnamefont {Perfetti}},\ }\href
  {https://doi.org/10.1103/physrevb.88.075120} {\bibfield  {journal} {\bibinfo
  {journal} {Physical Review B}\ }\textbf {\bibinfo {volume} {88}},\ \bibinfo
  {pages} {075120} (\bibinfo {year} {2013})}\BibitemShut {NoStop}%
\bibitem [{\citenamefont {Hase}\ \emph {et~al.}(1998)\citenamefont {Hase},
  \citenamefont {Mizoguchi}, \citenamefont {Harima}, \citenamefont {ichi
  Nakashima},\ and\ \citenamefont {Sakai}}]{Hase1998}%
  \BibitemOpen
  \bibfield  {author} {\bibinfo {author} {\bibfnamefont {M.}~\bibnamefont
  {Hase}}, \bibinfo {author} {\bibfnamefont {K.}~\bibnamefont {Mizoguchi}},
  \bibinfo {author} {\bibfnamefont {H.}~\bibnamefont {Harima}}, \bibinfo
  {author} {\bibfnamefont {S.}~\bibnamefont {ichi Nakashima}},\ and\ \bibinfo
  {author} {\bibfnamefont {K.}~\bibnamefont {Sakai}},\ }\href
  {https://doi.org/10.1103/physrevb.58.5448} {\bibfield  {journal} {\bibinfo
  {journal} {Physical Review B}\ }\textbf {\bibinfo {volume} {58}},\ \bibinfo
  {pages} {5448} (\bibinfo {year} {1998})}\BibitemShut {NoStop}%
\bibitem [{\citenamefont {G{\'{e}}neaux}\ \emph {et~al.}(2021)\citenamefont
  {G{\'{e}}neaux}, \citenamefont {Timrov}, \citenamefont {Kaplan},
  \citenamefont {Ross}, \citenamefont {Kraus},\ and\ \citenamefont
  {Leone}}]{Geneaux2021}%
  \BibitemOpen
  \bibfield  {author} {\bibinfo {author} {\bibfnamefont {R.}~\bibnamefont
  {G{\'{e}}neaux}}, \bibinfo {author} {\bibfnamefont {I.}~\bibnamefont
  {Timrov}}, \bibinfo {author} {\bibfnamefont {C.~J.}\ \bibnamefont {Kaplan}},
  \bibinfo {author} {\bibfnamefont {A.~D.}\ \bibnamefont {Ross}}, \bibinfo
  {author} {\bibfnamefont {P.~M.}\ \bibnamefont {Kraus}},\ and\ \bibinfo
  {author} {\bibfnamefont {S.~R.}\ \bibnamefont {Leone}},\ }\href
  {https://doi.org/10.1103/physrevresearch.3.033210} {\bibfield  {journal}
  {\bibinfo  {journal} {Physical Review Research}\ }\textbf {\bibinfo {volume}
  {3}},\ \bibinfo {pages} {033210} (\bibinfo {year} {2021})}\BibitemShut
  {NoStop}%
\bibitem [{\citenamefont {König}\ \emph {et~al.}(2021)\citenamefont {König},
  \citenamefont {Greer},\ and\ \citenamefont {Fahy}}]{Koenig2021}%
  \BibitemOpen
  \bibfield  {author} {\bibinfo {author} {\bibfnamefont {C.}~\bibnamefont
  {König}}, \bibinfo {author} {\bibfnamefont {J.~C.}\ \bibnamefont {Greer}},\
  and\ \bibinfo {author} {\bibfnamefont {S.}~\bibnamefont {Fahy}},\ }\href
  {https://doi.org/10.1103/physrevb.104.035127} {\bibfield  {journal} {\bibinfo
   {journal} {Physical Review B}\ }\textbf {\bibinfo {volume} {104}},\ \bibinfo
  {pages} {035127} (\bibinfo {year} {2021})}\BibitemShut {NoStop}%
\bibitem [{\citenamefont {{Anisimov}}\ \emph {et~al.}(1974)\citenamefont
  {{Anisimov}}, \citenamefont {{Kapeliovich}},\ and\ \citenamefont
  {{Perel'Man}}}]{Anisimov1974}%
  \BibitemOpen
  \bibfield  {author} {\bibinfo {author} {\bibfnamefont {S.~I.}\ \bibnamefont
  {{Anisimov}}}, \bibinfo {author} {\bibfnamefont {B.~L.}\ \bibnamefont
  {{Kapeliovich}}},\ and\ \bibinfo {author} {\bibfnamefont {T.~L.}\
  \bibnamefont {{Perel'Man}}},\ }\href@noop {} {\bibfield  {journal} {\bibinfo
  {journal} {Soviet Journal of Experimental and Theoretical Physics}\ }\textbf
  {\bibinfo {volume} {39}},\ \bibinfo {pages} {375} (\bibinfo {year}
  {1974})}\BibitemShut {NoStop}%
\bibitem [{\citenamefont {Stevens}\ \emph {et~al.}(2002)\citenamefont
  {Stevens}, \citenamefont {Kuhl},\ and\ \citenamefont {Merlin}}]{Stevens2002}%
  \BibitemOpen
  \bibfield  {author} {\bibinfo {author} {\bibfnamefont {T.~E.}\ \bibnamefont
  {Stevens}}, \bibinfo {author} {\bibfnamefont {J.}~\bibnamefont {Kuhl}},\ and\
  \bibinfo {author} {\bibfnamefont {R.}~\bibnamefont {Merlin}},\ }\href
  {https://doi.org/10.1103/physrevb.65.144304} {\bibfield  {journal} {\bibinfo
  {journal} {Physical Review B}\ }\textbf {\bibinfo {volume} {65}},\ \bibinfo
  {pages} {144304} (\bibinfo {year} {2002})}\BibitemShut {NoStop}%
\bibitem [{\citenamefont {Cardona}(1975)}]{Cardona1975}%
  \BibitemOpen
  \bibinfo {editor} {\bibfnamefont {M.}~\bibnamefont {Cardona}},\ ed.,\ \href
  {https://doi.org/10.1007/978-3-540-37568-5} {\emph {\bibinfo {title} {Light
  Scattering in Solids}}}\ (\bibinfo  {publisher} {Springer Berlin
  Heidelberg},\ \bibinfo {year} {1975})\BibitemShut {NoStop}%
\end{thebibliography}%
	
\end{document}